\documentclass[onefignum,onetabnum]{siamonline250211}

\usepackage{amsfonts}
\usepackage{graphicx}
\usepackage{epstopdf}
\usepackage{algorithmic}
\usepackage{algorithm}
\usepackage{subfig}
\usepackage{comment}
\usepackage{mathtools}
\usepackage{tabularx}
\usepackage{array}
\usepackage{multirow}
\usepackage{tablefootnote}

\ifpdf
  \DeclareGraphicsExtensions{.eps,.pdf,.png,.jpg}
\else
  \DeclareGraphicsExtensions{.eps}
\fi

\usepackage{enumitem}
\setlist[enumerate]{leftmargin=.5in}
\setlist[itemize]{leftmargin=.5in}

\newsiamremark{remark}{Remark}
\newsiamremark{hypothesis}{Hypothesis}
\crefname{hypothesis}{Hypothesis}{Hypotheses}
\newsiamthm{claim}{Claim}
\newsiamremark{fact}{Fact}
\crefname{fact}{Fact}{Facts}

\headers{Embedding networks with the random walk first return time distribution}{V. Thapar, R. Lambiotte, and G. T. Cantwell}

\title{Embedding networks with the random walk first return time distribution}

\author{Vedanta Thapar\thanks{The Mathematical Institute, University of Oxford, Oxford OX2 6GG, United Kingdom 
  (\email{vedanta.thapar@maths.ox.ac.uk}, \email{renaud.lambiotte@maths.ox.ac.uk}).}
\and Renaud Lambiotte\footnotemark[1]
\and George T. Cantwell\thanks{Department of Engineering, University of Cambridge, Cambridge CB2 1PZ, United Kingdom
  (\email{gtc31@cam.ac.uk}).}}

\usepackage{amsopn}

\ifpdf
\hypersetup{
  pdftitle={Embedding networks with the FRTD},
  pdfauthor={V. Thapar, R. Lambiotte, and G. T. Cantwell},
  colorlinks=true
}
\fi

\begin{document}

\maketitle

\begin{abstract}
We propose the first return time distribution (FRTD) of a random walk as an interpretable and mathematically grounded node embedding. 
The FRTD assigns a probability mass function to each node, allowing us to define a distance between any pair of nodes using standard metrics for discrete distributions. We present several arguments to motivate the FRTD embedding.
First, we show that FRTDs are strictly more informative than eigenvalue spectra, yet insufficient for complete graph identification, thus placing FRTD equivalence between cospectrality and isomorphism.
Second, we argue that FRTD equivalence between nodes captures structural similarity.
Third, we empirically demonstrate that the FRTD embedding outperforms manually designed graph metrics in network alignment tasks.
Finally, we show that random networks that approximately match the FRTD of a desired target also preserve other salient features.
Together these results demonstrate the FRTD as a simple and mathematically principled embedding for complex networks.
\end{abstract}

\section{Introduction} \label{Intro}
Network science attempts to provide the tools to understand the structure and function of high-dimensional interconnected systems \cite{newman2003structure, newman2018networks,lambiotte2021modularity}.
The most basic form of network analysis is descriptive---understanding networks or nodes with simple summary statistics.
A problem of particular interest is \emph{network embedding}, namely, representing nodes as vectors in an appropriate space so that structural properties of the network are encapsulated \cite{fouss2016algorithms}. 
Network embeddings play a crucial role in a wide range of data science tasks including community detection and node classification \cite{proximity, hamilton2017representation, cai2018comprehensive}.

There are many methods for embedding networks.
One strategy is to represent each node using a handful of heuristic metrics such as degree, clustering coefficient, or centrality measures \cite{Costa01012007}.
More advanced machine learning methods capture richer global patterns; many rely on spectral information derived from the graph Laplacian or the properties of random walk processes \cite{perozzi2014deepwalk, grover2016node2vec, ribeiro2017struc2vec, henderson2012rolx}.
However, machine-learning embedding methods are not without issues.
First, learning-based approaches may be stochastic and hence do not necessarily identify automorphically equivalent nodes with the same embedding, even though they are indistinguishable from a structural perspective.
Second, they may suffer from poor interpretability due to opaque neural network based transformations of graph features.

In this work we propose the first return time distribution (FRTD) of a random walk as an embedding for network nodes.
We focus on undirected networks but the approach is easily extended to weighted and directed networks.
We find that the FRTD captures meaningful structural information in a simple representation, mapping each node to a one-dimensional discrete probability distribution.

Several motivations underlie this study of random walk return embeddings.
Prior work has shown that random walk-based embeddings can enhance the performance of graph neural networks \cite{zhang2018retgk, dwivedi2021graph} and that diffusion kernels yield effective graph embeddings \cite{graphwave, Multidynam}.
Additionally, random walk returns are invariant under node relabeling and, although not invertible (that is, the FRTD does not uniquely determine the graph), we believe the corresponding preimage is typically very small and almost always unique.
We further demonstrate that FRTD information is strictly richer than the eigenspectrum alone, situating it between graph cospectrality and full isomorphism.

In contrast to prior work, we advocate specifically for the \emph{first} return time distribution, which to our knowledge is not currently used.
The shift is important because the FRTD is a distribution and hence normalized to sum to unity. 
On the other hand, return time probabilities (i.e., not the \emph{first} returns) converge to a constant that depends only on node degree.
The consequence is that the distance between (non-first) return time probabilities is only finite for nodes of the same degree.

We define notions for FRTD-equivalence and FRTD-distance between both nodes and graphs.
To illustrate the practical utility of the FRTD as an embedding, we present three applications:
role extraction, alignment, and randomization.

First, we consider the problem of grouping nodes into roles.
In a graph, nodes are considered to have similar roles if they are somehow structurally similar to each other \cite{browet2014algorithms, cason2012role}. 
We see that the FRTD embedding places nodes close together if they play a similar structural role in a network.
In contrast, embeddings based on diffusion-type dynamics are designed such that proximity in the embedding corresponds to proximity in the network \cite{proximity}; clustering nodes according to their first return time distributions yields partitions distinct from traditional community structure.

Second, we experimentally assess graph alignment algorithms that leverage FRTDs:
nodes in two networks are identified with each other so as to minimize the FRTD distances.
We show that the FRTD can be used to inform and improve existing alignment algorithms such as FUGAL \cite{bommakanti2024fugal}, which otherwise rely on heuristic network embeddings.

Finally, we look at how FRTDs can be used to randomize a network so that structural features are approximately preserved.
The configuration model, for example, provides a simple method for randomizing a graph while preserving the degree sequence.
However, it fails to preserve the vast majority of interesting features
such as triad counts, spectral features, or centrality rankings \cite{newman2003structure}.
The upshot is that this method does not create realistic looking networks.
It is an open problem to find a general purpose model that can generate networks that look realistic.
We show how the FRTD embedding can be used to define a model that interpolates between the $G(n,m)$ random graph ensemble and a model that preserves significant higher-order structure.

Taken together our arguments provide converging lines of evidence that the FRTD is a principled and useful method for embedding networks.

\section{Theory} \label{theory}

\subsection{Notation and background}
We consider undirected (but possibly weighted) graphs, $G = (V, E)$ ($|V| = n,\ |E| = m$).
Without loss of generality, we assume $V = \{1, 2, \dots n\}$, and hence the graph is defined by an $n \times n$ symmetric adjacency (or weight) matrix
\begin{equation}
    A_{ij} = \begin{cases}
        w_{ij} & \text{if}\ (i, j) \in E \\
        0 & \text{otherwise}
    \end{cases},
\end{equation}
where $w_{ij} = 1$ for an unweighted network. 
We will additionally assume weights to be positive, i.e., $w_{ij} > 0$.
The degree of node~$i$ is the sum over its edges, $d_i = \sum_{j} A_{ij}$, and the diagonal degree matrix has elements
\begin{equation}
	D_{ii} = d_i = \sum_{j} A_{ij}.
\end{equation}

A random walk $(u_0, u_1, \dots)$ that is currently at node $i$ will move to neighbor $j$ with probability
\begin{equation}
	P[u_{t+1}=j \vert u_{t}=i] = \frac{A_{ij}}{d_i},
\end{equation}
or in matrix notation the transition matrix is $\mathbf{D}^{-1} \mathbf{A}$.
To study the behavior of random walks we will pay particular attention to the symmetric normalized adjacency matrix
\begin{equation}
	\mathbf{X} = \mathbf{D}^{-1/2} \mathbf{A} \mathbf{D}^{-1/2} = 
	\mathbf{D}^{1/2} \left( \mathbf{D}^{-1} \mathbf{A} \right) \mathbf{D}^{-1/2}
	.
\end{equation}
Note that $\mathbf{X}$ is real and symmetric and hence has real eigenvalues.
Further, because $\mathbf{D}^{-1} \mathbf{A}$ is a stochastic matrix, and because $\mathbf{X}$ is related to this through a similarity transformation, the eigenvalues of $\mathbf{X}$ are in $[-1, 1]$.

Writing $\{ \lambda_\alpha \}_{\alpha=1}^{N}$ for the $N$ distinct eigenvalues of $\mathbf{X}$ with multiplicities $\{ N_\alpha \}_{\alpha=1}^{N}$ and corresponding orthonormal eigenvectors
$\{ \boldsymbol{\psi}_{\alpha \beta} \}_{\beta=1}^{N_\alpha}$,
then the \emph{spectral weight} of node~$i$ at $\lambda$ is defined to be
\begin{equation}
	w_i(\lambda) = \sum_{\alpha=1}^{N} \sum_{\beta=1}^{N_\alpha} \left( \psi_{\alpha \beta i} \right)^2 \delta_{\lambda, \lambda_\alpha},
	\label{eq:spec_weight}
\end{equation}
where $\delta_{x, y}=1$ if $x=y$ and $0$ otherwise.
Because the eigenvectors are orthonormal the spectral weight is normalized:
\begin{equation}
	\sum_{\alpha=1}^{N} w_i(\lambda_\alpha) = 1,
	\label{eq:spec_normalization}
\end{equation}
and
\begin{equation}
	\sum_{i} w_i(\lambda_\alpha) = N_\alpha .
	\label{eq:spec_normalization2}
\end{equation}

\subsection{First return time distributions}
\begin{figure}
    \centering
    \includegraphics[width=0.7\linewidth]{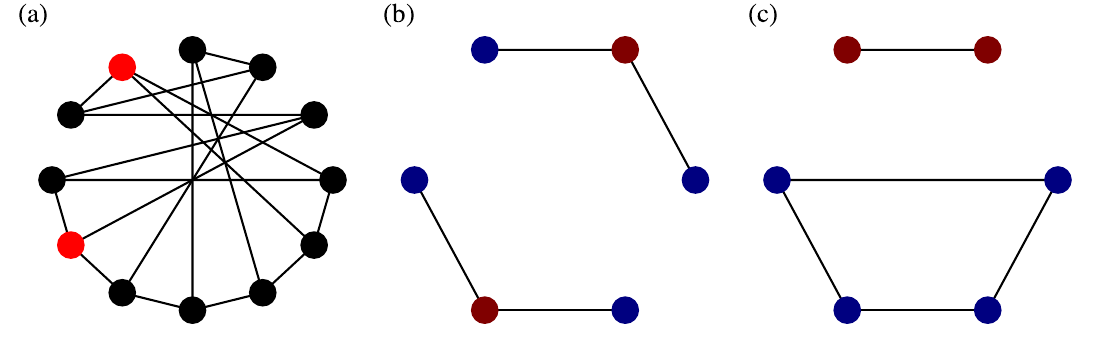}
	\caption{In panel (a) we show the Frucht graph \cite{Frucht_1949}. The two red nodes are not automorphically equivalent but share the same first return time distributions. 
	In (b) and (c) we give the smallest example of two non-isomorphic graphs that are FRTD equivalent, the nodes are colored according to FRTD equivalence across the two graphs.}
    \label{fig:FRTD_non_iso}
\end{figure}

Denote by $f_i(t)$ the probability that a random walk that starts at node~$i$ will return to $i$ for the first time after exactly $t$ steps,
\begin{equation}
	f_i(t) = P[u_t = i, u_s \neq i \text{ for } 0<s<t | u_0 = i].
\end{equation}
This is the \emph{first return time distribution} (FRTD) for node~$i$.
Our proposal is to interpret $f_i(t)$ as an embedding for node~$i$, i.e., to associate node~$i$ with the vector $\boldsymbol{f}_i$.
We immediately see the embedding is normalized since, on any finite network, a random walk must eventually return and so $\sum_t f_i(t) = 1$.
From the FRTD we define the following notions of equivalence.
\begin{definition}[FRTD equivalent nodes]
Let $f^{(G_1)}_i(t)$ be the FRTD of node $i$ in network $G_1$, and 
$f^{(G_2)}_j(t)$ be the FRTD of node $j$ in network $G_2$. 
We say these nodes are FRTD-equivalent if $f_i^{(G_1)}(t) = f_j^{(G_2)}(t)$ for all $t$.
\end{definition}
\begin{definition}[FRTD equivalent graphs]
Two graphs $G_1 = (V_1, E_1)$ and $G_2 = (V_2, E_2)$ are FRTD equivalent if there is a bijection between the nodes $\pi : V_1 \to V_2$ such that each $i \in V_1$ is FRTD equivalent to $\pi(i) \in V_2$.
\end{definition}
Note that when we compare two nodes in a single graph, then automorphically equivalent nodes (those in the same orbit) must be FRTD equivalent.
The converse is not true, as shown in the counter-example of Fig.~\ref{fig:FRTD_non_iso}(a).
Automorphic equivalence is the existence of a permutation that maps node $i$ to $j$, while leaving adjacency unchanged.
On the other hand, FRTD equivalence is any orthogonal transform (which may or may not be a permutation) that maps node $i$ to $j$, while leaving adjacency unchanged, see Theorem~\ref{thrm:FRTD_orthogonal}.

The early terms and moments of the FRTD have clear structural interpretations. By Kac's lemma the expected first return time to node $i$ on a connected graph is
\begin{equation}
    \sum_{t=0}^{\infty} t f_i(t) = \frac{1}{\pi_i} = \frac{2m}{d_i},
\end{equation}
where $\pi_i$ is the stationary probability of the walk being at $i$. Thus, the mean first return time encodes the stationary probability and, consequently, the degree of the node. For a simple undirected graph, the probability of returning after two steps is
\begin{equation}
    f_i(2) = \frac{1}{d_i}\sum_{j}\frac{A_{ij}}{d_j},
\end{equation}
which depends on both the degree of \(i\) and the degrees of its neighbors. Similarly, the probability of returning for the first time after three steps is
\begin{equation}
    f_i(3) =\sum_{j,k} \frac{A_{ij}A_{jk}A_{ki}}{d_i d_j d_k},
\end{equation}
which is a degree-weighted measure of the triangles containing $i$. Higher-order terms encode progressively longer closed structures while excluding trajectories that have already returned, providing an interpretable explanation for why the FRTD captures structural information beyond conventional local node statistics.

To further analyze FRTDs, it is convenient to work with their generating functions.
The first return times at node~$i$ are generated by
\begin{equation}
  F_i(z) = \sum_{t=0}^{\infty} z^t f_i(t),
\end{equation}
which has the following useful representation.
\begin{lemma}
For an undirected graph with normalized adjacency matrix $\mathbf{X}$,
let $w_i(\lambda)$ be the spectral weights as per Eq.~\eqref{eq:spec_weight}.
Then, 
\begin{equation}
	F_i(z) = 1 + \left(\sum_{\alpha=1}^{N} \frac{w_i(\lambda_\alpha)}{\lambda_\alpha z - 1} \right)^{-1}.
  \label{eq:spectral_frtd}
\end{equation}
\label{lemma:spectral_frtd}
\end{lemma}
\begin{proof}
First return times at node~$i$ are generated by $F_i(z)$ and $k$th return times are generated by $F_i(z)^k$.
Returns of any order are thus generated by $\sum_{k=0}^{\infty} F_i(z)^k = \left( 1 - F_i(z) \right)^{-1}$.
At the same time, the probability of a return of any order at time $t$ is
${\left(\mathbf{D}^{-1}\mathbf{A} \right)^t}_{ii} = \left( \mathbf{X}^t \right)_{ii}$ and hence
\begin{equation}
	\left( 1 - F_i(z) \right)^{-1} = \sum_{t=0}^{\infty} z^t \left(\mathbf{X}^t \right)_{ii} = \sum_{t=0}^{\infty}  z^t \sum_{\alpha=1}^{N} \lambda_{\alpha}^t \sum_{\beta=1}^{N_\alpha} (\psi_{\alpha \beta i})^2 = \sum_{\alpha=1}^{N} \frac{w_i(\lambda_\alpha) }{1 - \lambda_\alpha z},
\end{equation}
from which Eq.~\eqref{eq:spectral_frtd} immediately follows.
\end{proof}

With this representation for the generating function $F_i(z)$ we quickly find the following results.
\begin{theorem}
\label{thrm:FRTD_equiv}
For nodes~$i \in G_1$ and $j \in G_2$, let $w_i^{(G_1)}(\lambda)$ and $w_j^{(G_2)}(\lambda)$ be the respective spectral weight functions.
Nodes $i \in G_1$ and $j \in G_2$ are FRTD-equivalent if and only if $w_i^{(G_1)}(\lambda) = w_j^{(G_2)}(\lambda)$ for all $\lambda$.
\end{theorem}
\begin{proof}
One direction is immediate: if the spectral weight functions match then by the spectral representation of Lemma~\ref{lemma:spectral_frtd} the generating functions $F_i^{(G_1)}(z) = F_j^{(G_2)}(z)$ and so the distributions are identical.

Conversely, if $f_i^{(G_1)}(t) = f_j^{(G_2)}(t)$ for all $t$ then $F^{(G_1)}_i(z) = F^{(G_2)}_j(z)$.
Additionally, note that the function $(F^{(G_1)}_i(z)-1)^{-1}$ has simple poles at exactly those $1/\lambda^{(G_1)}_\alpha$ for which both $\lambda^{(G_1)}_\alpha \neq 0$ and $w_i^{(G_1)}(\lambda^{(G_1)}_\alpha) \neq 0$, with residue $w_i^{(G_1)}(\lambda^{(G_1)}_\alpha)/\lambda^{(G_1)}_\alpha$.
The equivalent statement for $(F^{(G_2)}_j(z)-1)^{-1}$, together with the equality of poles and residues, establishes that $w_i^{(G_1)}(\lambda) = w_j^{(G_2)}(\lambda)$ for all $\lambda \neq 0$.
Finally, the normalization of Eq.~\eqref{eq:spec_normalization} fixes $w_i^{(G_1)}(0) = w_j^{(G_2)}(0)$.
\end{proof}

If $G_1$ and $G_2$ are cospectral, i.e., they have the same eigenvalues and multiplicities, FRTD-equivalence of nodes $i \in G_1$ and $j \in G_2$ is equivalent to $w_i^{(G_1)}(\lambda_\alpha)=w_j^{(G_2)}(\lambda_\alpha)$ for all $\alpha$.
In particular:
\begin{remark}
Two nodes $i$ and $j$ in the same graph are FRTD-equivalent if and only if $w_i(\lambda_\alpha)=w_j(\lambda_\alpha)$ for all $\alpha$.
\label{remark:frtd_same_graph}
\end{remark}
Additionally, FRTD-equivalence between graphs is at least as strong as cospectrality:
\begin{corollary}
If $G_1$ and $G_2$ are FRTD-equivalent then their normalized adjacency matrices are cospectral.
\end{corollary}
\begin{proof}
FRTD-equivalence and Theorem~\ref{thrm:FRTD_equiv} imply $w_i^{(G_1)}(\lambda) = w_{\pi(i)}^{(G_2)}(\lambda)$ for all $i \in G_1$ and $\lambda$, and some appropriate permutation $\pi(i)$.
Writing $N^{(G)}(\lambda)$ for the multiplicity of $\lambda$ in the spectrum of $G$, Eq.~\eqref{eq:spec_normalization2} gives 
\begin{equation}
      N^{(G_1)}(\lambda) = \sum_i w_i^{(G_1)}(\lambda) = \sum_i w_{\pi(i)}^{(G_2)}(\lambda) = N^{(G_2)}(\lambda)
\end{equation}
and so the eigenvalues and multiplicities match.
\end{proof}
In fact, FRTD-equivalence is strictly stronger than cospectrality.
To see this, note that the $4$-cycle $C_4$ and the $3$-star $S_3$ both have eigenvalues $(-1, 0, 0, 1)$ but are clearly not FRTD-equivalent.
At the same time, FRTD-equivalence is weaker than isomorphism as the counter-example in Fig.~\ref{fig:FRTD_non_iso} shows (or see Appendix~\ref{app:noniso_FRTD_equiv} for a larger counter example, but where the graph is connected).
Hence, FRTD-equivalence sits between isomorphism and cospectrality.
Although it has not been proved, non-isomorphic cospectral graphs in the (non-normalized) adjacency matrix appear to be exceedingly rare \cite{haemers, Wang03042025}.
For this reason we conjecture that almost all graphs have no FRTD-equivalent twins that are not isomorphic.

Finally, we note that FRTD-equivalence is a weakening of automorphic equivalence in a particular sense.
Nodes $i,j \in G$ are automorphically equivalent if there is a permutation matrix $\pi$ with $\pi \mathbf{X} = \mathbf{X} \pi $ and $\pi \boldsymbol{e}_i = \boldsymbol{e}_j$, where $\boldsymbol{e}_i$ is the unit vector for node~$i$.
FRTD equivalence relaxes this to any orthogonal matrix, as seen by the following theorem.
\begin{theorem}
	Nodes $i, j \in G$ with normalized adjacency $\mathbf{X}$ are FRTD equivalent if and only if there exists an orthogonal $\mathbf{Q}$ with
\begin{equation*}
	\mathbf{Q} \mathbf{X}  = \mathbf{X} \mathbf{Q} \quad\quad \text{and} \quad\quad \mathbf{Q} \boldsymbol{e}_i = \boldsymbol{e}_j.
\end{equation*}
\label{thrm:FRTD_orthogonal}
\end{theorem}
\begin{proof}
Let the projection onto eigenspace $\alpha$ be denoted
\begin{equation}
	\mathbf{\Psi}_\alpha = \sum_{\beta=1}^{N_\alpha} \boldsymbol{\psi}_{\alpha \beta} \boldsymbol{\psi}_{\alpha \beta}^{T}.
\end{equation}

If $\mathbf{Q}$ commutes with $\mathbf{X} = \sum_\alpha \lambda_\alpha \boldsymbol{\Psi}_\alpha$, then $\mathbf{Q}$ commutes with each $\mathbf{\Psi}_\alpha$ individually.  
Noting that 
$w_i(\lambda_\alpha) = \vert \mathbf{\Psi}_\alpha \boldsymbol{e}_i \vert^2$
we get
\begin{equation}
	w_j(\lambda_\alpha) = \vert \mathbf{\Psi}_\alpha \boldsymbol{e}_j \vert^2 = \vert \mathbf{\Psi}_\alpha \mathbf{Q} \boldsymbol{e}_i\vert^2 = \vert \mathbf{Q} \mathbf{\Psi}_\alpha \boldsymbol{e}_i \vert^2 = \vert \mathbf{\Psi}_\alpha \boldsymbol{e}_i \vert^2 = w_i(\lambda_\alpha)
\end{equation}
where the penultimate equality is due to $\mathbf{Q}$ not changing norms.
This is FRTD equivalence by Remark~\ref{remark:frtd_same_graph}.

For the other direction, if $w_i(\lambda_\alpha) = w_j(\lambda_\alpha)$ then the vectors $\mathbf{\Psi}_\alpha \boldsymbol{e}_i$ and $\mathbf{\Psi}_\alpha \boldsymbol{e}_j$ are equal in magnitude.
Either they are also equal in direction, or they can be mapped onto each other by a reflection in the eigenspace,  i.e., by an orthogonal $\mathbf{Q}_{\alpha}$.
In the first case, set $\mathbf{Q}_\alpha = \mathbf{I}$, in the second set $\mathbf{Q}_\alpha = \mathbf{I} - 2\boldsymbol{q}\boldsymbol{q}^T / \vert q \vert^2 $ where $\boldsymbol{q} = \mathbf{\Psi}_\alpha (\boldsymbol{e}_i - \boldsymbol{e}_j)$.
In either case we now have an orthogonal $\mathbf{Q}_\alpha$ such that
\begin{enumerate}[label=(\roman*)]
	\item $\mathbf{Q}_\alpha \mathbf{\Psi}_\alpha = \mathbf{\Psi}_\alpha \mathbf{Q}_\alpha$, and
	\item $\mathbf{Q}_\alpha \mathbf{\Psi}_\alpha \boldsymbol{e}_i = \mathbf{\Psi}_\alpha \boldsymbol{e}_j$.
\end{enumerate}
Now set 
\begin{equation}
	\mathbf{Q} = \sum_{\alpha=1}^{N} \mathbf{Q}_\alpha \mathbf{\Psi}_\alpha
\end{equation}
and note this orthogonal $\mathbf{Q}$ commutes with $\mathbf{X}$ by (i), and by (ii) we have
\begin{equation}
	\mathbf{Q} \boldsymbol{e}_i = \sum_\alpha \mathbf{Q}_\alpha \mathbf{\Psi}_\alpha \boldsymbol{e}_i = \sum_\alpha \mathbf{\Psi}_\alpha \boldsymbol{e}_j = \boldsymbol{e}_j.
\end{equation}
\end{proof}

\subsection{Node and graph distances}
While exact FRTD equivalence may be rare, our central contention is that if two nodes (resp. graphs) have similar but not identical FRTDs, those nodes (resp. graphs) are meaningfully similar to one another.
To complete the interpretation of $f_i(t)$ as an embedding we must additionally choose a distance metric, $d(\boldsymbol{f}_i, \boldsymbol{f}_j)$.
Any distance between probability mass functions is a possible candidate for $d$, though in this work we consider the total variation distance
\begin{equation}
  d(\boldsymbol{f}_i, \boldsymbol{f}_j) = \frac{1}{2} \sum_{t=0}^{\infty} \left\vert  f_i(t) - f_j(t) \right\vert.
\end{equation}
The total variation is a bounded distance metric between distributions with a maximum value of $1$.
Note, however, it does not define a metric between nodes since $d(\boldsymbol{f}_i, \boldsymbol{f}_j) = 0$ does not imply $i=j$.

The node-level measure can be aggregated to define a distance between graphs,
\begin{equation}
	d(G_1, G_2) = \frac{1}{n} \sum_{i=1}^{n} d(\boldsymbol{f}_i^{(G_1)}, \boldsymbol{f}_i^{(G_2)})
\end{equation}
and note that if $d(G_1, G_2)=0$ then $G_1$ and $G_2$ are FRTD-equivalent. In cases where the labelling is unknown the distance may be defined as the minimum labelled distance over all vertex bijections, i.e., $\min_{\pi:V_1\rightarrow V_2} \frac{1}{n} \sum_{i\in V_1} d( f_i^{(G_1)}, f_{\pi(i)}^{(G_2)} )$. A minimum can be found efficiently with the Hungarian algorithm \cite{kuhn1955hungarian}.

Again, while exact FRTD equality may be exceedingly rare we contend that small values for $d(\boldsymbol{f}_i, \boldsymbol{f}_j)$ or $d(G_1, G_2)$ imply that the nodes or graphs are meaningfully similar, and for this reason $\boldsymbol{f}_i$ is a sensible embedding of the nodes.
In the remaining sections of this paper we present a sequence of experiments to provide evidence for our claim that FRTD embeddings encode meaningful properties of graphs.

\section{Empirical study of FRTDs} \label{empirical}
If the FRTD is a good embedding then it should be useful for downstream tasks.
We now present experiments using the FRTD for role extraction, graph alignment, and graph randomization.

\subsection{Role extraction}
Identifying meaningful groupings of nodes is a standard task for summarizing and analyzing networks.
Much of this work focuses on community structure \cite{FORTUNATO20161}, and often assortative community structure where nodes are assumed to connect preferentially to nodes in their same group.
However, we believe that the FRTD similarity does not speak to this issue.
For example, in the stochastic block model---the main generative model for community structure---nodes that are in different groups may well have the same FRTD \cite{PhysRevE.111.064306}. This is further illustrated in Sec. \ref{synthetic_graphs}.

We propose that the FRTD does not distinguish nodes based on community structure but rather roles.
Intuitively, two nodes have the same role if they are structurally similar, regardless of their mutual proximity or community membership \cite{rossi2014role}. There are a variety of approaches to role extraction such as generalized block-modelling \cite{doreian2005generalized}, clustering of node self-similarity measures \cite{cason2012role, browet2014algorithms}, and methods relying on the properties of the graph Laplacian, for example GraphWave \cite{graphwave} which uses diffusion patterns (governed by the heat kernel) to learn structural node embeddings. The machine learning literature has also addressed the problem,
for example RolX \cite{henderson2012rolx}
which uses node features (such as degree, triangle count, etc.) to cluster roles, and struc2vec \cite{ribeiro2017struc2vec} which constructs a multilayer graph of structural similarities between nodes on which it samples random walks to learn role-based embeddings.
A review of approaches can be found in \cite{proximity}. 

To illustrate the role information captured by the FRTD we consider a sequence of numerical experiments on synthetic graphs with planted structural roles and a set of real world networks with metadata describing relevant `roles'. As a point of comparison three other role discovery methods are tested: RolX \cite{henderson2012rolx}, struc2vec \cite{ribeiro2017struc2vec}, and GraphWave \cite{graphwave}. To describe the influence of structure beyond the first hop neighbourhood, we also consider node degree as a `trivial' benchmark role embedding. To illustrate the distinction between role assignment and community detection tasks, we also test a spectral embedding constructed with the leading eigenvectors of $\mathbf{X}$. Such embeddings have been widely used for community detection tasks \cite{von_luxburg_tutorial_2007}.

In Fig.~\ref{fig:barbell} we display a simple example with easily interpretable clusters.
The FRTD embedding is compared to other role-based methods as well as to node2vec \cite{grover2016node2vec}, a proximity based embedding method.

\subsection{Results} \label{sec:role extraction}

To convert node embeddings into roles, we apply a standard $k$-nearest neighbors classifier.
Performance is assessed by 5-fold stratified cross-validation, and results are reported by the mean macro F1-score, which weights all roles equally regardless of size. 
We run experiments on a range of synthetic and real graphs.

\newlength{\figw}
\setlength{\figw}{4.75in} 
\begin{figure}
    \centering
    \subfloat[Barbell graph]{\includegraphics[width=0.66\figw]{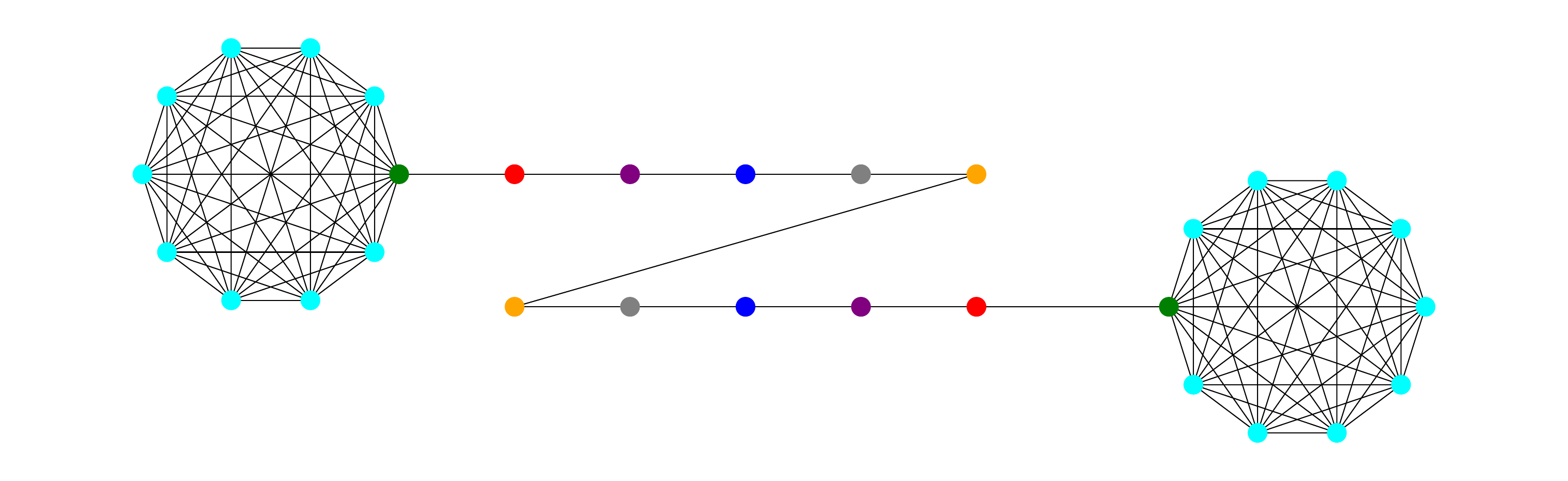}}
    \subfloat[node2vec \cite{grover2016node2vec} (proximity based \cite{proximity})]{\includegraphics[width=0.33\figw]{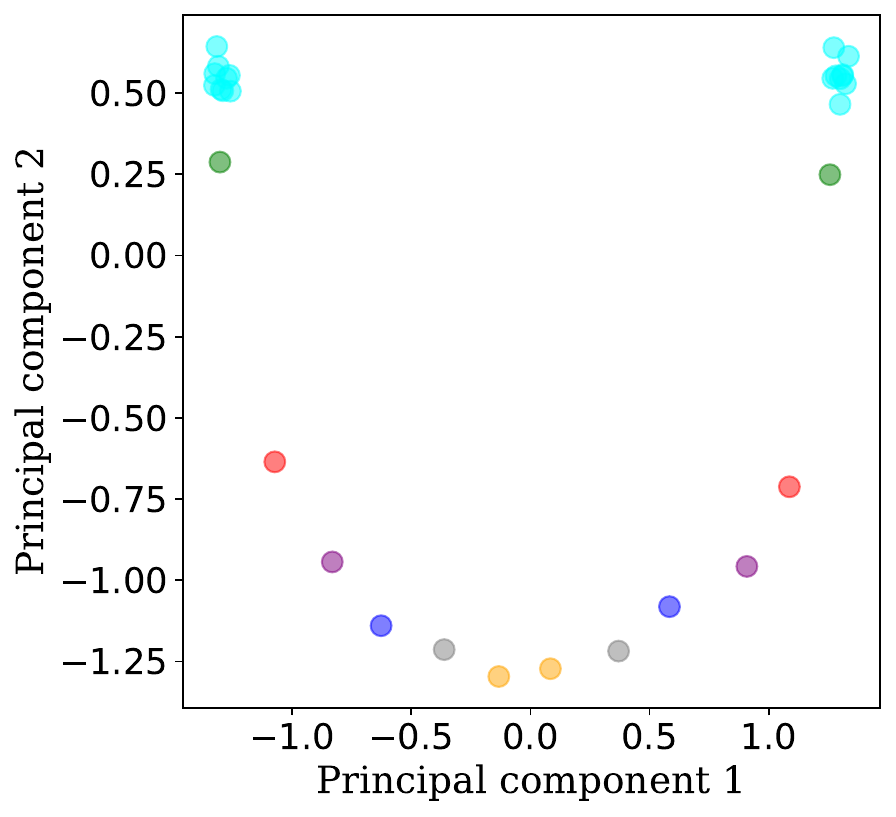}} \\
    \subfloat[struc2vec \cite{ribeiro2017struc2vec}]{\includegraphics[width=0.33\figw]{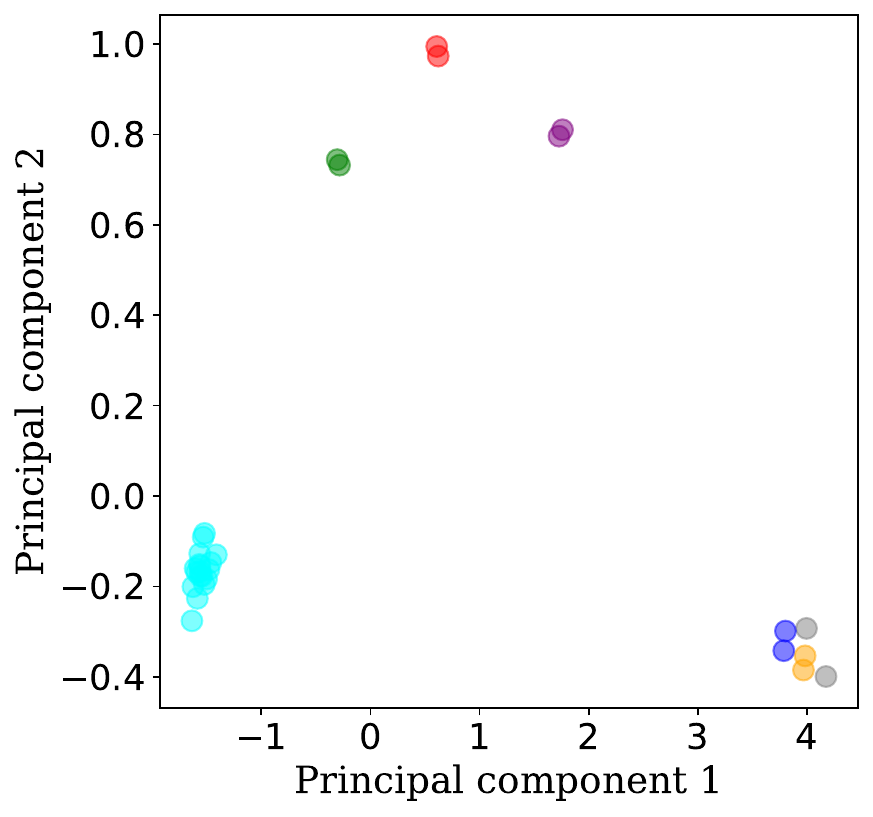}}
    \subfloat[GraphWave \cite{graphwave}]{\includegraphics[width=0.33\figw]{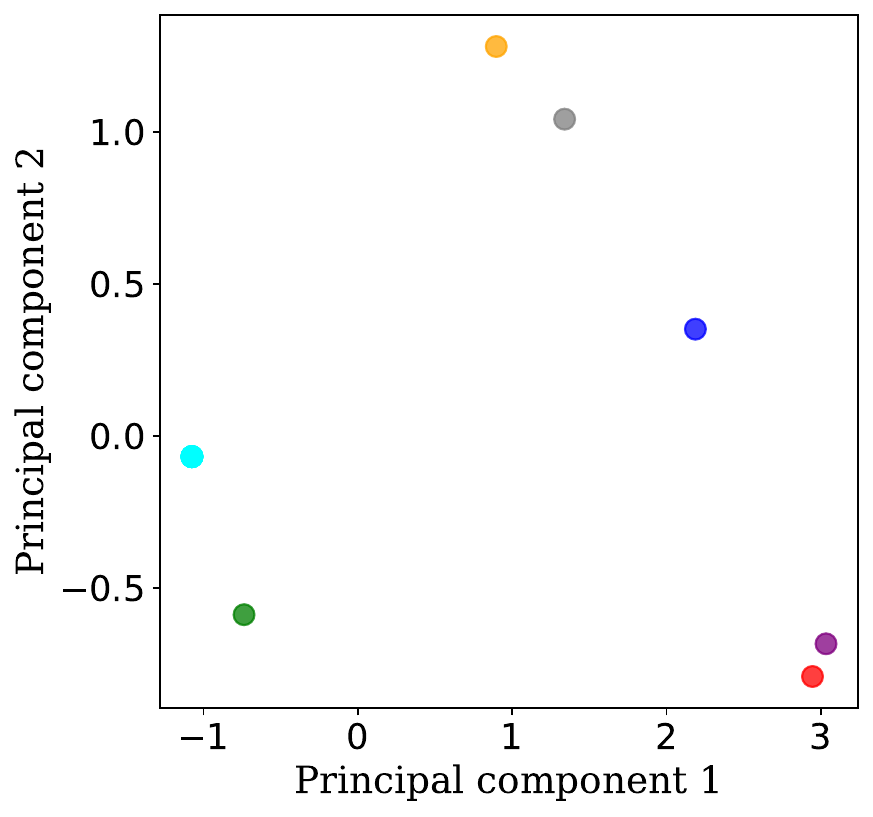}}
    \subfloat[FRTD]{\includegraphics[width=0.33\figw]{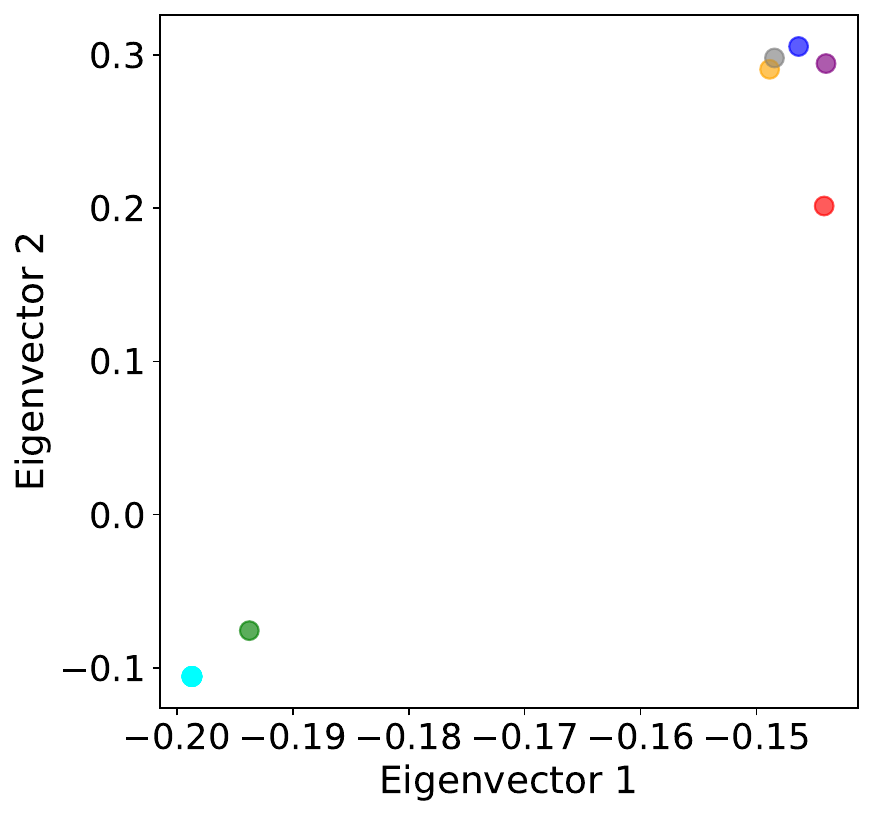}}
    \caption{Principal component visualization of different embedding methods for the Barbell graph. For the FRTD, leading eigenvectors of the similarity matrix with elements $s_{ij} = e^{-d(\boldsymbol{f}_i, \boldsymbol{f}_j)}$ are used in place of principal components.
	Embeddings are calculated with the default parameters for all the algorithms. 
	node2vec is designed to place nodes that are close to each other in the graph close in the embedding space.
	The other methods are structural (role) based \cite{proximity}.
	Notably struc2vec does not tend to identify automorphically equivalent nodes with the same embeddings.
	In contrast, GraphWave assigns structurally equivalent nodes to arbitrarily close embeddings (with appropriate hyper-parameter selection) and FRTD embeddings are always exactly equivalent for automorphic nodes.}
    \label{fig:barbell}
\end{figure}

\subsubsection{Synthetic graphs} \label{synthetic_graphs}
Intuitively, node roles should correspond to different nodes doing different things.
In the context of a generative model, this should mean that different nodes connect to the graph according to different rules.
(Again, roles should not be confused with communities---the same role may exist in different communities, and each community may contain multiple roles.)

To generate synthetic graphs with two planted roles, we generate networks by combining two different random graph models.
In particular, we look at networks constructed by combining:
\begin{itemize}
    \item {\bf Random geometric \cite{penrose2003random} and Erdős–Rényi graphs, (RG+ER)}. A graph on $n=n_1 + n_2$ nodes with average degree $ k$ is constructed by first sampling a random geometric graph of $n_1$ nodes with radius $r=\sqrt{\frac{k}{\pi (n-1)}}$ and then attaching $n_2$ nodes independently at random to each of the $n-1$ nodes with probability $p = \frac{k}{n-1}$. These parameters ensure that the roles have the same mean degree, so degree alone cannot distinguish them.
    \item {\bf Preferential attachment and duplication divergence (PA+DD)}. A graph on $n$ nodes is constructed sequentially, starting with a clique on $l+1$ nodes. New nodes are added and choose where to connect. With probability $1/2$ the new node attaches by the preferential attachment mechanism, i.e., it selects $l$ existing nodes with probability proportional to their degree. Otherwise, the node attaches using the duplication divergence rule, i.e., it randomly chooses a single node and attaches to it. Then, it additionally connects to each of the chosen node's neighbors, independently and with probability $q$.
    \item {\bf Stochastic block model}. To illustrate the distinction between roles and communities, we also consider the stochastic block model (SBM) in the assortative regime as a baseline generative model where the node labels correspond instead to communities and not roles.  Role detection methods are expected to perform poorly.
\end{itemize}

\begin{figure}
    \centering
    \includegraphics[width=\linewidth]{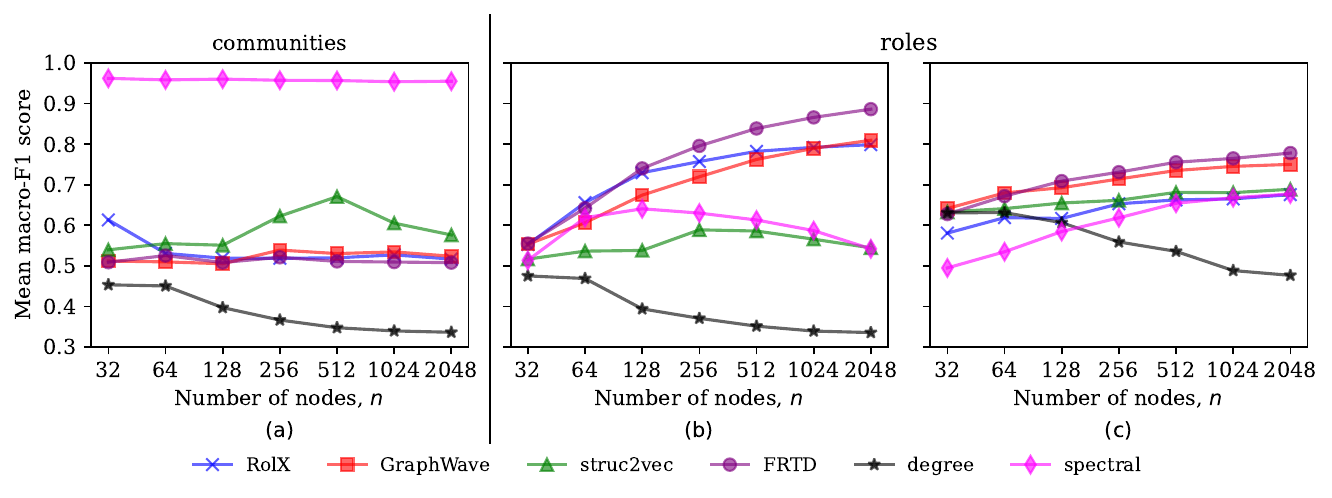}
    \caption{Results of 5-fold stratified cross validation on synthetic graphs with planted roles with increasing graph size. Panel (a) displays the performance on a two community symmetric stochastic block model in the assortative sparse regime with within block connection probability $p_{\text{in}} = \frac{16}{n}$, and between block $p_{\text{out}}=\frac{4}{n}$. Here the `ground truth' labels are the block memberships. Panels (b)-(c) display the results for mixed generative models where the node labels correspond to roles---(b) displays the results for RG+ER graphs, here we fix an average degree of $k=8$. Similarly panel (c) presents the results for PA+DD graphs,  with fixed $q=\frac{l-1}{2l-1}$ and $l=6$ in order to match the expected degrees of the two growth mechanisms.  All experiments were run with the default hyperparameters for each algorithm (see Appendix \ref{role hyperparameters} for details).} 
    \label{fig:synthetic_results}
\end{figure}

In Fig.~\ref{fig:synthetic_results} we show the performance of the embedding methods with varying network size. While spectral embedding performs well on the community detection task, its ability to distinguish roles is limited. On the mixed generative models all role-finding methods perform reasonably and significantly outperform node degree in isolation. In contrast on the SBM they remain around the random guessing baseline. The FRTD consistently performs better than other methods on role extraction tasks and has competitive runtime.

\subsubsection{Real graphs}

To simulate real application, we now apply the methods to empirical data, with metadata treated as a proxy for the node's structural role\footnote{While metadata labels do not constitute ground-truth roles, they provide an external benchmark for evaluating whether the embeddings capture meaningful structural distinctions between nodes.}.

In Table~\ref{real-world networks} we report the performance of the benchmark embedding methods on 8 real world graphs from different domains, with the associated definition of a `role' in the data. No single method is the best across all networks; the FRTD performs competitively across all datasets and has the highest macro F1-score on approximately half of them. 
Hence, although the FRTD is not \emph{designed} to pick out roles, it reliably performs on par or better than methods that were designed specifically for this task.

\begin{table}[]
\resizebox{\textwidth}{!}{%
\begin{tabular}{|cccc|ccccc|}
\hline
\multicolumn{4}{|c|}{\textbf{dataset}}                                                                                                              & \multicolumn{5}{c|}{\textbf{\begin{tabular}[c]{@{}c@{}}mean macro-F1 score\\ time (s)\end{tabular}}}                                                                                                                                                                                                                                                                                                     \\ \hline
\multicolumn{1}{|c|}{\textbf{name}}   & \multicolumn{1}{c|}{\textbf{n}} & \multicolumn{1}{c|}{\textbf{type}}  & \textbf{notion of role}             & \multicolumn{1}{c|}{\textbf{degree}}                                      & \multicolumn{1}{c|}{\textbf{RolX}}                                         & \multicolumn{1}{c|}{\textbf{graphwave}}                                              & \multicolumn{1}{c|}{\textbf{struc2vec}}                                             & \textbf{FRTD}                                                 \\ \hline
\multicolumn{1}{|c|}{adjnoun \cite{newman2006finding, tiago_p_peixoto_2020_7839981}}         & \multicolumn{1}{c|}{112}        & \multicolumn{1}{c|}{word adjacency} & word type (noun/adjective)          & \multicolumn{1}{c|}{\begin{tabular}[c]{@{}c@{}}0.535\\ 0.01\end{tabular}} & \multicolumn{1}{c|}{\begin{tabular}[c]{@{}c@{}}0.654\\ 0.07\end{tabular}}  & \multicolumn{1}{c|}{\begin{tabular}[c]{@{}c@{}}0.562\\ 0.02\end{tabular}}            & \multicolumn{1}{c|}{\textbf{\begin{tabular}[c]{@{}c@{}}0.668\\ 1.89\end{tabular}}}  & \begin{tabular}[c]{@{}c@{}}0.599\\ 0.04\end{tabular}          \\ \cline{5-9} 
\multicolumn{1}{|c|}{celegans \cite{cookelegans}}        & \multicolumn{1}{c|}{453}        & \multicolumn{1}{c|}{brain}          & neuron type                         & \multicolumn{1}{c|}{\begin{tabular}[c]{@{}c@{}}0.232\\ 0.01\end{tabular}} & \multicolumn{1}{c|}{\begin{tabular}[c]{@{}c@{}}0.548\\ 0.51\end{tabular}}  & \multicolumn{1}{c|}{\begin{tabular}[c]{@{}c@{}}0.658\\ 0.34\end{tabular}}            & \multicolumn{1}{c|}{\begin{tabular}[c]{@{}c@{}}0.613\\ 31.95\end{tabular}}          & \textbf{\begin{tabular}[c]{@{}c@{}}0.722\\ 0.51\end{tabular}} \\ \cline{5-9} 
\multicolumn{1}{|c|}{webkb-cornell \cite{webkb, webkb-link}}   & \multicolumn{1}{c|}{183}        & \multicolumn{1}{c|}{internet}       & webpage type (student/faculty/etc.) & \multicolumn{1}{c|}{\begin{tabular}[c]{@{}c@{}}0.170\\ 0.01\end{tabular}} & \multicolumn{1}{c|}{\begin{tabular}[c]{@{}c@{}}0.251\\ 0.11\end{tabular}}  & \multicolumn{1}{c|}{\begin{tabular}[c]{@{}c@{}}0.258\\ 0.07\end{tabular}}            & \multicolumn{1}{c|}{\begin{tabular}[c]{@{}c@{}}0.244\\ 2.52\end{tabular}}           & \textbf{\begin{tabular}[c]{@{}c@{}}0.301\\ 0.09\end{tabular}} \\ \cline{5-9} 
\multicolumn{1}{|c|}{webkb-texas \cite{webkb, webkb-link}}     & \multicolumn{1}{c|}{182}        & \multicolumn{1}{c|}{internet}       & webpage type (student/faculty/etc.) & \multicolumn{1}{c|}{\begin{tabular}[c]{@{}c@{}}0.242\\ 0.01\end{tabular}} & \multicolumn{1}{c|}{\begin{tabular}[c]{@{}c@{}}0.531\\ 0.11\end{tabular}}  & \multicolumn{1}{c|}{\begin{tabular}[c]{@{}c@{}}0.383\\ 0.06\end{tabular}}            & \multicolumn{1}{c|}{\textbf{\begin{tabular}[c]{@{}c@{}}0.546\\ 2.34\end{tabular}}}  & \begin{tabular}[c]{@{}c@{}}0.438\\ 0.08\end{tabular}          \\ \cline{5-9} 
\multicolumn{1}{|c|}{webkb-wisconsin \cite{webkb, webkb-link}} & \multicolumn{1}{c|}{251}        & \multicolumn{1}{c|}{internet}       & webpage type (student/faculty/etc.) & \multicolumn{1}{c|}{\begin{tabular}[c]{@{}c@{}}0.165\\ 0.01\end{tabular}} & \multicolumn{1}{c|}{\begin{tabular}[c]{@{}c@{}}0.331\\ 0.18\end{tabular}}  & \multicolumn{1}{c|}{\begin{tabular}[c]{@{}c@{}}0.292\\ 0.11\end{tabular}}            & \multicolumn{1}{c|}{\begin{tabular}[c]{@{}c@{}}0.259\\ 4.03\end{tabular}}           & \textbf{\begin{tabular}[c]{@{}c@{}}0.372\\ 0.15\end{tabular}} \\ \cline{5-9} 
\multicolumn{1}{|c|}{airports-brazil \cite{ribeiro2017struc2vec}} & \multicolumn{1}{c|}{131}        & \multicolumn{1}{c|}{transportation} & airport type                        & \multicolumn{1}{c|}{\begin{tabular}[c]{@{}c@{}}0.744\\ 0.01\end{tabular}} & \multicolumn{1}{c|}{\begin{tabular}[c]{@{}c@{}}0.619\\ 0.07\end{tabular}}  & \multicolumn{1}{c|}{\begin{tabular}[c]{@{}c@{}}0.736\\ 0.04\end{tabular}}            & \multicolumn{1}{c|}{\begin{tabular}[c]{@{}c@{}}0.758\\ 3.84\end{tabular}}           & \textbf{\begin{tabular}[c]{@{}c@{}}0.792\\ 0.06\end{tabular}} \\ \cline{5-9} 
\multicolumn{1}{|c|}{airports-europe \cite{ribeiro2017struc2vec}} & \multicolumn{1}{c|}{399}        & \multicolumn{1}{c|}{transportation} & airport type                        & \multicolumn{1}{c|}{\begin{tabular}[c]{@{}c@{}}0.552\\ 0.01\end{tabular}} & \multicolumn{1}{c|}{\begin{tabular}[c]{@{}c@{}}0.521\\ 0.45\end{tabular}}  & \multicolumn{1}{c|}{\begin{tabular}[c]{@{}c@{}}0.581\\ 0.28\end{tabular}}            & \multicolumn{1}{c|}{\textbf{\begin{tabular}[c]{@{}c@{}}0.597\\ 33.04\end{tabular}}} & \begin{tabular}[c]{@{}c@{}}0.568\\ 0.45\end{tabular}          \\ \cline{5-9} 
\multicolumn{1}{|c|}{openflights\footnotemark}     & \multicolumn{1}{c|}{3188}       & \multicolumn{1}{c|}{transportation} & airport type                        & \multicolumn{1}{c|}{\begin{tabular}[c]{@{}c@{}}0.549\\ 0.02\end{tabular}} & \multicolumn{1}{c|}{\begin{tabular}[c]{@{}c@{}}0.697\\ 20.07\end{tabular}} & \multicolumn{1}{c|}{\textbf{\begin{tabular}[c]{@{}c@{}}0.716\\ 143.72\end{tabular}}} & \multicolumn{1}{c|}{\begin{tabular}[c]{@{}c@{}}0.685\\ 592.54\end{tabular}}         & \begin{tabular}[c]{@{}c@{}}0.693\\ 27.05\end{tabular}         \\ \hline
\end{tabular}%
}
\caption{Results of 5-fold stratified cross validation on a set of 8 real world networks, results reported are the mean macro-f1 score and computation time, the best performing algorithm is labelled in bold for each dataset. All experiments were run with the default hyperparameters for each algorithm (see Appendix \ref{role hyperparameters} for details).}
\label{real-world networks}
\end{table}
\footnotetext{Upstream: \href{https://openflights.org/}{openflights.org}, retrieved from 
\href{https://networks.skewed.de/net/openflights}
{networks.skewed.de/net/openflights} \cite{tiago_p_peixoto_2020_7839981}. Airport sizes were taken from the OurAirports dataset(\href{https://ourairports.com/data/}{ourairports.com/data}).
}

\subsection{Alignment} \label{alignment}
We now turn to graph alignment.
Two graphs $G_1 = (V_1, E_1)$ and $G_2=(V_2, E_2)$ are assumed to be on the same set of nodes.
However, the identification of nodes in $V_1$ with those in $V_2$ is unknown and the task of graph alignment is to find this mapping.
Usually we seek a mapping between the nodes so that the (labeled) topology is very close between the two graphs.
We refer to Ref.~\cite{TANG20251} for a review.

To make this task a well-posed mathematical problem, we ask for a permutation $\pi$ such that the following (squared) Frobenius norm is minimized
\begin{equation}
	L(\pi) = || \mathbf{A}^{(1)} \pi - \pi \mathbf{A}^{(2)} ||_F^2,
  \label{eq:alignment}
\end{equation}
where $\mathbf{A}^{(1)}$ is the adjacency matrix of $G_1$, and likewise for $G_2$.
That is to say, we seek $\pi = \arg\min_{\pi'} L(\pi')$.
Clearly the $\min_{\pi'} L(\pi') = 0$ if and only if the graphs are isomorphic.
However, the optimization is well defined even when no zero solution exists, and hence alignment can be considered a relaxation of the isomorphism problem \cite{doi:10.1073/pnas.1401651112}.

Minimizing Eq.~\eqref{eq:alignment} is a quadratic assignment problem and is hence NP-hard.
In general, efficient algorithms will not find the global minimum but many fast heuristics have been developed, such as gradient descent methods (see Ref.~\cite{faq}).
One approach maps the problem to a linear sum assignment by defining a cost matrix $C_{ij}$ which assigns a cost to matching a node  $i \in V_1$ to $j \in V_2$, i.e., we instead look for the bijection $\pi : V_1 \to V_2$ that minimizes
\begin{equation}
    \sum_{i} C_{i, \pi(i)},
    \label{eq:linear_assignment}
\end{equation}
and the minimum can be found with $\mathcal{O}(n^3)$ operations by, e.g., the Hungarian algorithm \cite{kuhn1955hungarian}.
This cost $C_{ij}$ is typically assigned by defining a notion of distance between nodes.

The linear assignment of Eq.~\eqref{eq:linear_assignment} 
will not generally be equivalent to the full quadratic problem of Eq.~\eqref{eq:alignment}.
The quality of the solution will depend on how well the structure is captured by the entries of $\mathbf{C}$.
The most competitive approaches combine both and minimize
\begin{equation}
	|| \mathbf{A}^{(1)} \pi - \pi \mathbf{A}^{(2)} ||_F^2 + \mu \  \text{Tr} (\pi^T\mathbf{C}),
    \label{eq:QAP+LAP}
\end{equation}
which allows the matrix $\mathbf{C}$ to guide the heuristic solver for the quadratic assignment problem.
This is the approach of Feature-Fortified Unrestricted Graph Alignment (FUGAL) \cite{bommakanti2024fugal},
which 
defines $\mathbf{C}$ by a grab-bag of node features, such as degrees, neighbor degrees, and clustering coefficients.

Instead of reaching for a battery of hand selected features, we believe the FRTD captures many of the relevant structural properties.
For this reason we propose setting $C_{ij} = d(\boldsymbol{f}_i^{(G_1)}, \boldsymbol{f}_j^{(G_2)})$.
We insert this cost matrix directly into the FUGAL algorithm's approximate solver and henceforth refer to this as FUGAL-FRT.

To experimentally assess the method, we produce corrupted copies of graphs 
and attempt to recover the alignment.
We follow Ref.~\cite{skitsas2023comprehensive} and consider benchmarking examples as follows:
\begin{itemize}
    \item Input graph $G_1$ and initialize graph $G_2$ by copying $G_1$.
    \item Randomly remove $x \%$ of edges in $G_2$.
    \item Permute the node labels in $G_2$, with this permutation serving as the ``ground truth'' mapping.
\end{itemize}
We can repeat the process several times to produce a set of pairs of graphs $(G_1, G_2)$ with known alignments.
To measure the quality of a recovered permutation $\pi$ we compute the proportion of correctly aligned nodes.

\begin{figure}
    \centering
    \includegraphics[width=\linewidth]{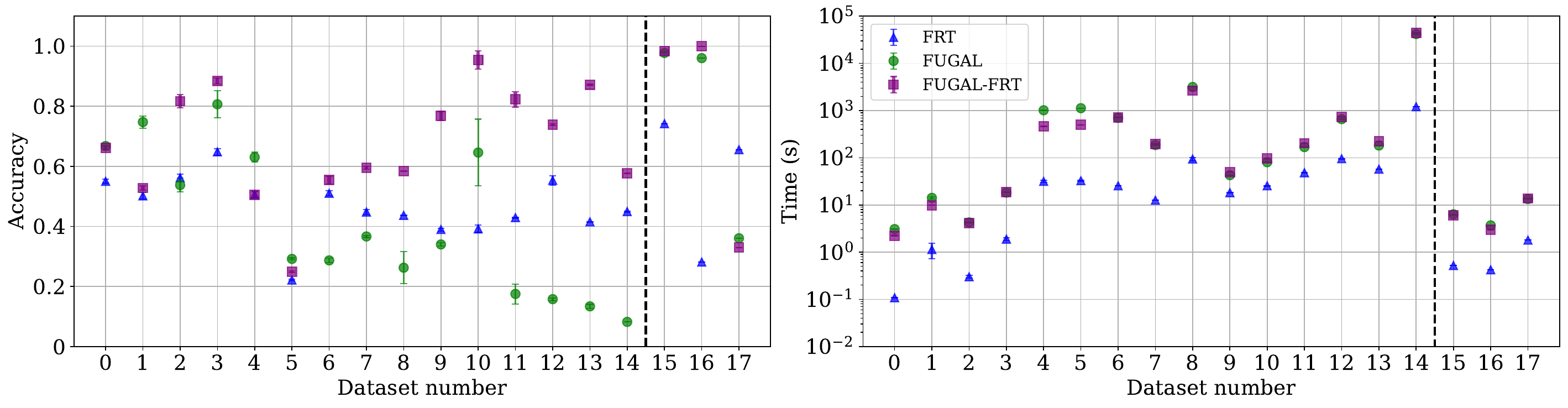}
    \caption{Performance of three alignment algorithms on a set of 18 undirected network datasets ordered by number of edges---see Appendix \ref{GAL datasets} for details. 
	For the first 15 graphs, the noisy copy is created by randomly removing $5\%$ of edges and permuting the nodes.
	For the last three networks, ``real'' noise is used (see Appendix \ref{GAL datasets} for details). Panel (a) reports the accuracy (defined as the proportion of correctly aligned nodes) while panel (b) displays the runtime for each algorithm.
	All experiments were run with the same hyperparameters ($\mu=1$) and no optimization of hyperparameters was considered.}
    \label{alignment_result1}
\end{figure}

We compare three different methods, namely FUGAL, FUGAL-FRT, and a pure linear assignment using only the FRTDs.
We test the approach on the networks previously used in Ref.~\cite{skitsas2023comprehensive}.
Figure~\ref{alignment_result1} shows the performance of the methods, 
we see that FRTD linear alignment is often less accurate than FUGAL and FUGAL-FRT, but it runs significantly faster in all cases.

More importantly, FUGAL-FRT generally performs better than the classic FUGAL, often considerably so.
This is notable because while FUGAL uses a variety of intuitively motivated metrics to embed nodes, FUGAL-FRT only uses the FRTD.
This provides further evidence that the FRTD contains a rich variety of structural information about nodes, and that being close in FRTD is a strong indication that nodes are structurally similar.

\subsection{Network randomization} \label{network_randomization}
Through the experiments of the previous two sections we have started to provide evidence that FRTDs are a meaningful embedding for nodes.
We now turn to the question of how well the FRTDs capture the whole structure of a network.
To this end, we consider whether the FRTDs form a suitable basis for network randomization.

We consider a random graph ensemble that interpolates between the Erdős–Rényi random graph $G(n,m)$, i.e., uniform over all graphs on $n$ nodes with $m$ edges, and a model that preserves the full FRTD.
Given a graph $G$, we generate a randomized version of it with the same number of edges.
Graph $G'$ is sampled with probability
\begin{equation}
    P_G(G') = \frac{e^{-\beta d(G, G')}}{Z_{G, \beta}},
	\label{eq:random_graph_model}
\end{equation}
where the parameter $\beta$ is an inverse temperature, $d$ is the FRTD distance between the graph $G'$ and the target graph $G$, and $Z_{G, \beta}$ is a normalizing constant that does not depend on $G'$.
We sample from the distribution with Markov chain Monte Carlo (MCMC) and, for computational efficiency, we estimate $d(G, G')$ by truncating the FRTDs at $K=14$.
Starting from an initial graph $G'$, we propose a new graph $G^*$ by a combination of two operations: with probability $0.4$ we select an edge at random and move it, else we perform a random degree preserving rewiring.
The modified graph is accepted according to the Metropolis criterion $\min\{1, e^{-\beta (d(G^{*}, G) - d(G', G))}\}$.

The inverse temperature, $\beta$, interpolates between uniformly random networks with the same degree sequence at $\beta = 0$, and graphs matching the target’s FRTDs as $\beta \to \infty$.
These distributions are difficult to sample from using MCMC \cite{park2004statistical, snijders2002markov, bhamidi2008mixing}, thus
to traverse high-probability regions separated by large barriers of low probability we employ parallel tempering: multiple chains are run at different $\beta$ and we periodically attempt to swap the temperatures of these chains \cite{PhysRevLett.57.2607, earl2005parallel}.

We study two examples: the Karate club network (34 nodes) \cite{zachary1977information}, and the \textit{C. elegans} connectome (297 nodes) \cite{white1986structure, watts1998collective}.

\subsubsection{Karate club} 
\begin{figure}
    \centering
    \includegraphics[width=\linewidth]{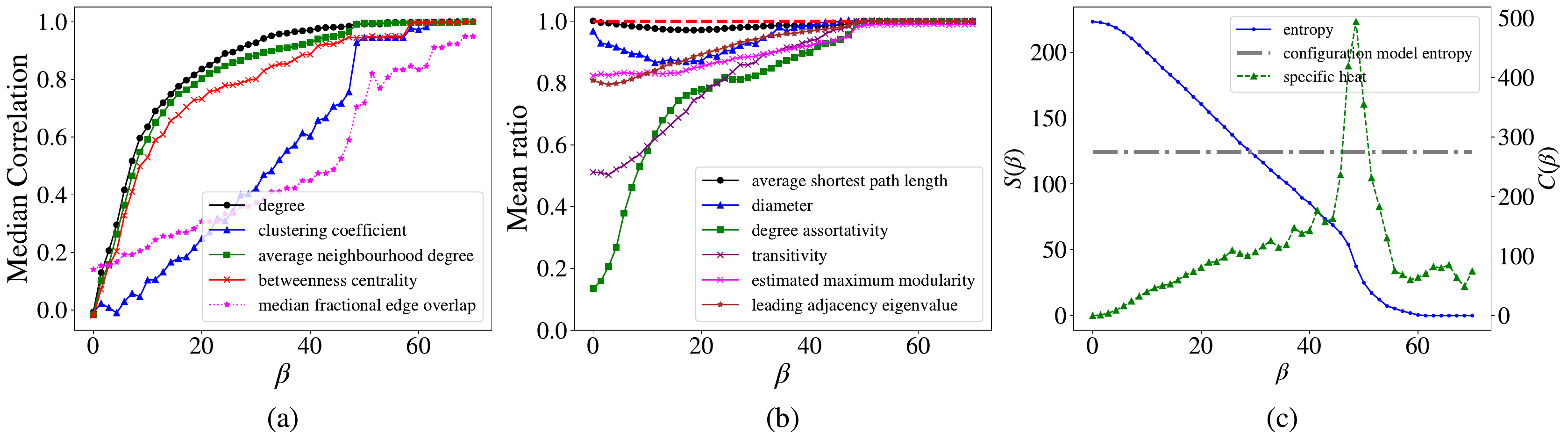}\\[-2.5ex]
    \includegraphics[width=\linewidth]{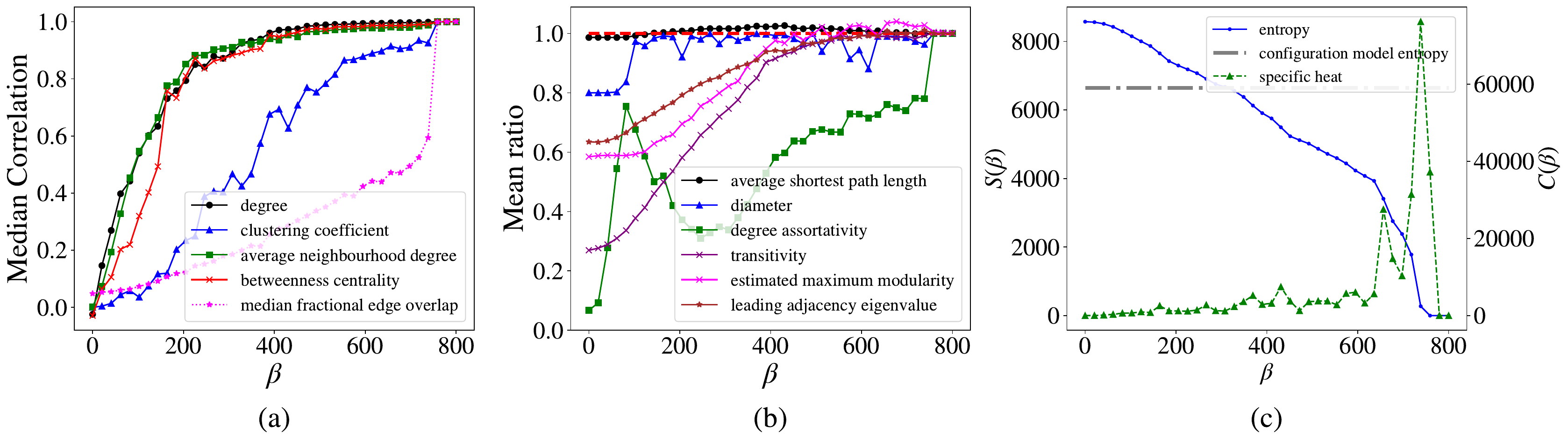}
    \caption{Graph statistics for the Karate club  (top row)
    and \textit{C. elegans} (bottom row).
    The statistics of the original network are compared with those obtained from $P_G(G')$ at different values of $\beta$.
    At $\beta=0$, we sample all graphs with the same number of edges with equal probability.
    Increasing $\beta$ biases the distribution towards networks with similar FRTDs to the target. Panel (a) shows the median Pearson correlation coefficient between various node level descriptors for nodes in $G'$ compared to $G$, i.e., half of the sampled graphs $G'$ have a Pearson correlation to $G$ which is larger than the reported median value. Panel (b) displays the average ratio of various global network descriptors for $G'$ to $G$, i.e., $E[s(G')/s(G)]$ for graph statistic $s$. Panel (c) shows the entropy and specific heat of the random graph ensemble. We also show the configuration model entropy, estimated from the approximation of McKay and Wormald  \cite{MCKAY1990565}.}
    \label{fig:stats_KC_CE}
\end{figure}

The Monte Carlo process utilizes $50$ chains with $\beta \in [0, 70]$ initialized at a random $G(n,m)$ graph. A burn-in period of $2\times10^6$ samples is employed after which we collect samples. At large $\beta$ the Karate club network is recovered exactly by the algorithm and as we reduce $\beta$ we sample random graphs which resemble the original graph in several respects.
In Fig.~\ref{fig:stats_KC_CE} we see that the samples preserve both local and global features, with the degree of preservation depending on the temperature.

As the properties of the random graphs approach the target, the entropy decreases.
The entropy (which roughly measures the logarithm of the number of ``typical'' graphs) and the specific heat (which measures how the expected value of $d(G',G)$ changes with $\beta$) both display a change in the region $\beta \approx 50$, indicating phase transition like behavior. 
Beyond this point, a large proportion of the samples are either isomorphic to the target or differ only by very small perturbations.
But, shortly before this point we see a region that still has a relatively large entropy but also preserves many properties of the graph.

\subsubsection{\textit{C. elegans}} For the \textit{ C. elegans} network
we see a similar picture. The bottom row of Fig.~\ref{fig:stats_KC_CE} displays the graph statistics obtained at different $\beta$.
As before, $\beta$ determines the interpolation between complete randomization of the edges and FRTD preserving randomization.
By increasing $\beta$ the various graph properties are better preserved.
Before a phase-transition like behavior, we see a region with non-zero entropy that still does a good job at preserving the statistics of the original graph.

To reiterate, the distribution of Eq.~\eqref{eq:random_graph_model} is defined entirely by the FRTD.
The fact that it preserves key statistics of the base graphs suggests that the FRTD itself encodes the relevant structural information.

\section{Discussion}
\label{sec:dis}

In this work we have argued that the first return time distribution of a random walk captures meaningful structural information and forms an effective graph embedding
method.
The approach is straightforwardly extended to weighted and directed networks---see Appendix~\ref{sec:numerical-comp}.
In addition to the theoretical results, the key evidence we provide in favor of FRTD embedding is a sequence of empirical experiments.
We have demonstrated that the FRTD can be used to (i) define (and hence discover) structural roles for the nodes in a network; (ii) identify nodes across corrupted versions of the same network; and (iii) define a sensible distance between graphs which can be used to generate random graphs that are still ``close'' to the original. In the role discovery experiments, the FRTD is observed to capture structural information associated with an intuitive notion of role or function as demonstrated by its performance on both synthetic and real world graphs relative to existing role discovery algorithms.

In the alignment experiments, we find that the FRTD embedding often outperforms a hand-crafted feature vector chosen for its intuitive appeal. This suggests that a single, well-motivated feature---the FRTD---can serve as a compact and effective summary of structural information.
However, as the role-extraction experiments show, the FRTD does not, in general, encode community structure. Rather than replacing community analysis, the FRTD provides complementary information.

In the randomization experiments we find that enforcing an exact match to the FRTDs can lead to a degenerate distribution, with essentially all probability mass on the original graph (consistent with the conjectures of Refs.~\cite{exponentialDS, haemers, Wang03042025}).
However, by relaxing this constraint slightly (i.e., decreasing $\beta$ in Eq.~\eqref{eq:random_graph_model}), we identify a region of non-zero entropy that still approximately preserves the salient structural features.

In summary, the FRTD is an informative embedding, 
normalized and invariant to node relabeling, and capable of capturing structural properties in a compact form.
Our results offer additional justification for the widespread use of random-walk–based statistics in machine learning, beyond the pragmatic observation that ``they appear to work in practice'' \cite{zhang2018retgk, dwivedi2021graph}.
From a practical standpoint, however, we recommend using the FRTD rather than return probabilities.

Open problems remain.
On a practical level, while it is easy to compute the FRTDs from a network it is far from trivial to reverse this mapping, even approximately.
The Markov chain we employ should technically achieve this for sufficiently large $\beta$, but the method is prohibitively slow for large networks.
An efficient method to generate networks with a given random walk structure could be of great interest to both theorists and practitioners. Finally, in this work we compute the FRTD exactly by repeated multiplication of the random walk matrix.
This method however scales as $\mathcal{O}(n^3K)$ in worst case (see Appendix \ref{sec:numerical-comp} for details).
To allow for greater scalability, the FRTD may also be estimated, for example by Monte Carlo \cite{Monte_Carlo_PageRank} or message passing \cite{PhysRevE.111.064306}.
These algorithms would allow for efficient and decentralized embedding, but at the cost that calculations would only be approximate.

\section*{Code and data availability}

The full code (along with publicly available links to data) to reproduce the results in this work can be found at \\ \href{https://github.com/Vedanta-T/RandomWalk_FRTD_embeddings.git}{https://github.com/Vedanta-T/RandomWalk\_FRTD\_embeddings.git}.

\section*{Funding}
VT was supported by Rhodes Trust and the Mathematical Institute, University of Oxford. RL acknowledges support from the EPSRC Grants EP/V013068/1, EP/V03474X/1, and EP/Y028872/1.

\section*{Acknowledgements}
We thank Erik Hormann for useful discussions.

\newpage

\appendix

\section{Non-isomorphic graphs with the same FRTDs}
\label{app:noniso_FRTD_equiv}
\begin{figure}
    \centering
    \includegraphics[width=0.5\linewidth]{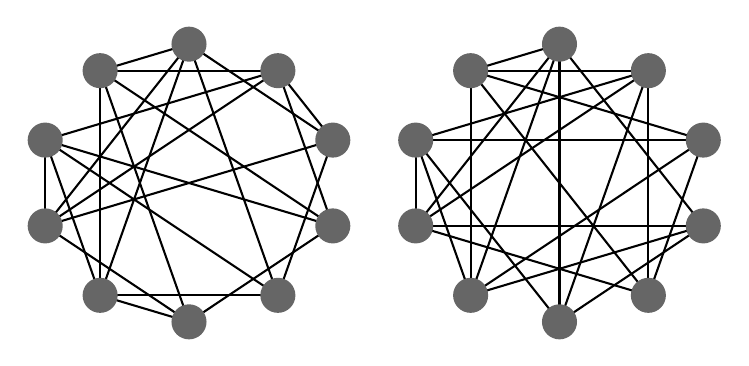}
	\caption{Two FRTD-equivalent graphs that are non-isomorphic. Nodes are ordered by their FRTD, so that nodes drawn in the same location for both graphs have identical FRTDs.}
    \label{fig:FRTD_simple_non_iso}
\end{figure}
The two graphs in Fig.~\ref{fig:FRTD_simple_non_iso} are FRTD-equivalent, although they are not isomorphic.
Further, they are both connected and have no degenerate eigenvalues, for either the adjacency or the normalized adjacency matrices.

These graphs in Fig.~\ref{fig:FRTD_simple_non_iso} are additionally asymmetric (the only automorphism is the identity), and each node has a unique FRTD.
In both drawings, the nodes are ordered by their (unique) FRTD so that the only permutation that preserves the FRTD between the two graphs is the identity.
The non-isomorphism can hence be seen trivially from the figure: any edge that exists in one graph but not the other is proof that the graphs are non-isomorphic.

\section{Computing FRTDs} \label{sec:numerical-comp}
The distribution $f_i(t)$ can be computed for all $i$ by matrix multiplication.
The following algorithm achieves this.
\begin{algorithm}[H]
    \caption{Algorithm to compute FRTDs}
    \textbf{Input}: $(n \times n)$ transition matrix $\mathbf{T} = \mathbf{D}^{-1}\mathbf{A}$ and maximum number of steps, $K$.
    \begin{algorithmic}[1]
    \STATE $\mathbf{P} \gets \text{Identity}(n)$
    \STATE $\mathbf{f} \gets \text{zeros}((n, K+1))$
    \FOR{$t$ in $1:K$}
    \STATE $\mathbf{P} \gets \mathbf{T}\mathbf{P}$
    \STATE $\mathbf{f}[:, t] \gets \text{diagonal}(\mathbf{P})$
    \STATE $\text{diagonal}(\mathbf{P}) = \mathbf{0}$
    \ENDFOR
    \STATE $\mathbf{f}[:,K+1] \gets \mathbf{1} - \mathbf{f}\, \mathbf{1}$
    \RETURN $\mathbf{f}$
    \end{algorithmic}    
\end{algorithm}
When $\mathbf{A}$ is sparse then the matrix multiplication $\mathbf{T}\mathbf{P}$ requires $\mathcal{O}(n^2)$ operations and hence the full algorithm runs in $\mathcal{O}(K n^2)$.
If $\mathbf{A}$ is not sparse, it will take $\mathcal{O}(K n^3)$ operations.
Note, in order to produce a normalized embedding the remaining probability mass is appended, i.e., $f[i, K+1] \gets 1 - \sum_{t=1}^K f[i, t]$.

\subsection{FRTDs for directed networks} \label{FRTDdirected}

The above definitions can be extended to construct FRTDs for directed networks by introducing ``teleportation'' to avoid getting stuck at sets of nodes with no outward edges.
At each step the random walker will move by an outgoing edge with probability $1-\alpha$ or alternatively with probability $\alpha$ it will ``teleport'' by jumping to any node in the network with equal probability.
The transition matrix is thus
\begin{equation}
	(1-\alpha) \mathbf{D}^{-1} \mathbf{A} + (\alpha/n) \mathbf{1}
	\label{eq:pagerank_transition}
\end{equation}
where $\mathbf{1}$ is a matrix of all ones.
Typically $\alpha$ is chosen to be $\alpha=0.15$, as originally suggested in the PageRank algorithm \cite{ilprints422}.

For an informative embedding that depends on both in-edges and out-edges, we use the FRTD of the transition matrix of Eq.~\eqref{eq:pagerank_transition}, together with the corresponding distribution on the edge-reversed graph, i.e. with $\mathbf{A}$ replaced by $\mathbf{A}^T$.

\subsection{Example} 
\begin{figure}
    \centering
    \includegraphics[width=\linewidth]{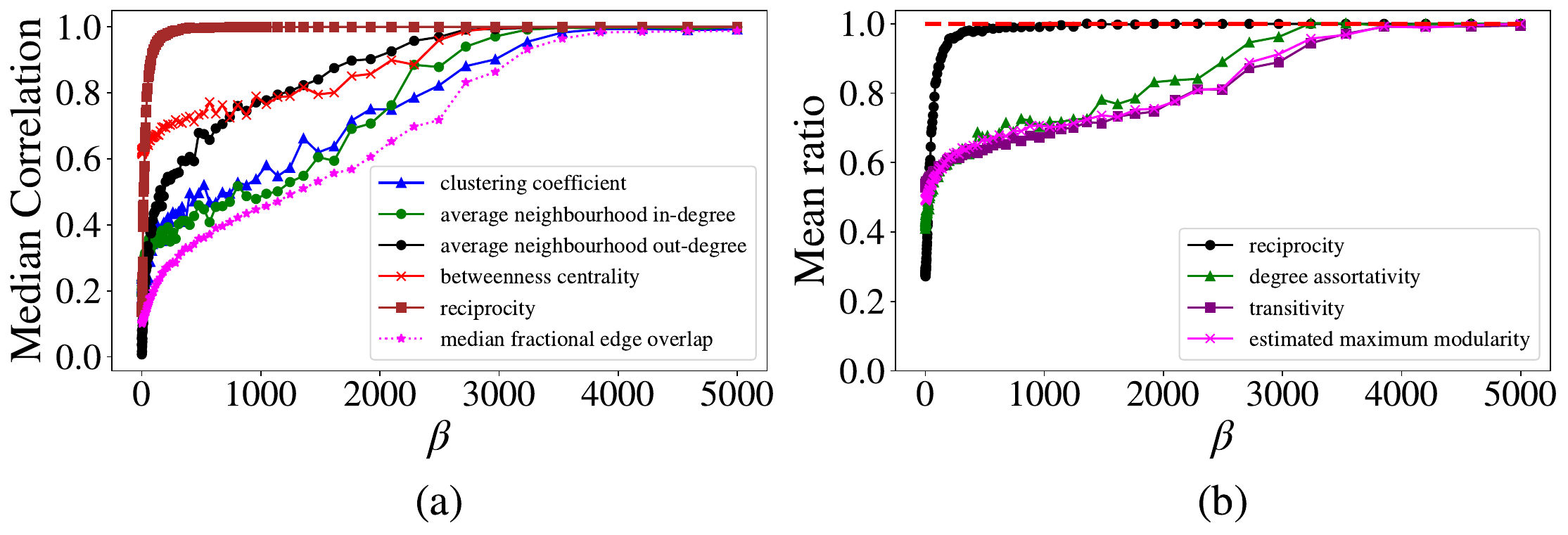}
    \caption{Graph statistics for the directed \textit{C. elegans} connectome compared with those obtained from $P_G(G')$ at different temperatures, note that $\beta=0$ corresponds to sampling all graphs with the same in- and out-degree sequence with equal probability. Panel (a) shows the median correlation between various node level descriptors of samples $G'$ and $G$, (b) displays the ratio of various global network descriptors for $G'$ vs $G$.}
    \label{fig:Celegans_directed}
\end{figure}

As an illustration of the above we study randomization of the directed \textit{C. elegans} graph (in the sense of Sec. \ref{network_randomization}). To improve the mixing time we fix the degree sequence of the random graphs and candidates are proposed only using the degree preserving rewiring method described in \cite{directedDP}.
One hundred chains with $\beta \in [0, 5000]$ are initialized at $G$ and a burn-in period of $2\times 10^5$ steps is used before collecting samples.
Fig. \ref{fig:Celegans_directed} shows the statistics of the samples.
This directed version of the model interpolates between the directed configuration model at $\beta=0$ and a model that preserves the full FRTD for both $\mathbf{A}$ and $\mathbf{A}^T$ as $\beta \to \infty$.

\section{Hyperparameters for role extraction benchmarks} \label{role hyperparameters}

Each algorithm considered in Sec. \ref{sec:role extraction} has its own set of hyperparameters, for our analysis we use the default values provided in the algorithm's respective original repositories. These values are summarized in Table \ref{tab:hyperparameters}

\begin{table}[t]
\centering
\begin{tabularx}{\textwidth}{|c|c|>{\raggedright\arraybackslash}X|}
\hline
\textbf{algorithm} &
\textbf{hyperparameters} &
\multicolumn{1}{|c|}{\textbf{description}} \\
\hline

\multirow{3}{*}{RolX \cite{henderson2012rolx}}
& \texttt{n\_roles} = 4
& predefined number of roles, also corresponds to dimensionality of embedding.
\\ \cline{2-3}

& \texttt{n\_iter} = 10
& number of ReFeX \cite{henderson2011refex} recursive feature extraction rounds.
\\ \cline{2-3}

& \texttt{max\_iter} = 2000
& maximum number of iterations for non-negative matrix factorization solver.
\\ \hline

\multirow{3}{*}{GraphWave \cite{graphwave}}
& \texttt{scales} = $(1.0,\,5.0)$
& the two heat-kernel diffusion scales used to compute wavelets.
\\ \cline{2-3}

& \texttt{n\_samples} = 50
& number of points at which the characteristic function of each wavelet is sampled.
\\ \cline{2-3}

& \texttt{t\_max} = 100
& upper bound of the time sampling range for the characteristic function used to generate embeddings.
\\ \hline

\multirow{5}{*}{struc2vec \cite{ribeiro2017struc2vec}}
& \texttt{walk\_length} = 40
& length of each random walk on the multilayer graph.
\\ \cline{2-3}

& \texttt{num\_walks} = 10
& number of random walks sampled.
\\ \cline{2-3}

& \texttt{window} = 10
& length of skip-gram context window for learning embeddings.
\\ \cline{2-3}

& \texttt{iter} = 5
& number of epochs for training word2vec.
\\ \cline{2-3}

& \texttt{opt3\_num\_layers}
& the maximum number of layers used for option 3 as provided in \cite{ribeiro2017struc2vec}; set to diameter of graph.
\\ \hline

FRTD
& \texttt{K} = 50
& depth of FRTD embedding.
\\ \hline
spectral
& \texttt{k} = 5
& number of leading eigenvectors considered.
\\ \hline

\end{tabularx}
\caption{List and description of hyperparameters used for role extraction benchmarking experiments in Sec.~\ref{sec:role extraction}. Detailed descriptions of each of these parameters can be found in the respective papers and repositories linked therein.}
\label{tab:hyperparameters}
\end{table}

\section{Index of datasets for graph alignment} \label{GAL datasets} We follow Ref.~\cite{skitsas2023comprehensive}, which reviews graph alignment methods using a standard dataset of 18 networks of varying sizes and domains. 
Table \ref{tab:alignmentdata} provides their details along with an index corresponding to the $x$ axis labels in Fig.~\ref{alignment_result1}. 

\begin{table}
\centering
\resizebox{\textwidth}{!}{%
\begin{tabular}{|ccccc|ccc|}
\hline
\multicolumn{5}{|c|}{\textbf{dataset}} &
  \multicolumn{3}{c|}{\textbf{\begin{tabular}[c]{@{}c@{}}accuracy (average $\pm$ standard deviation)\\ time in seconds (average $\pm$ standard deviation)\end{tabular}}} \\ \hline
\multicolumn{1}{|c|}{\textbf{index}} &
  \multicolumn{1}{c|}{\textbf{name}} &
  \multicolumn{1}{c|}{\textbf{$\boldsymbol{n}$}} &
  \multicolumn{1}{c|}{\textbf{$\boldsymbol{m}$}} &
  \textbf{type} &
  \multicolumn{1}{c|}{\textbf{FRT}} &
  \multicolumn{1}{c|}{\textbf{FUGAL}} &
  \textbf{FUGAL-FRT} \\ \hline
\multicolumn{1}{|c|}{\textbf{0}} &
  \multicolumn{1}{c|}{ca-netscience \cite{newman2006finding}} &
  \multicolumn{1}{c|}{379} &
  \multicolumn{1}{c|}{914} &
  collaboration &
  \multicolumn{1}{c|}{\begin{tabular}[c]{@{}c@{}}$0.550 \pm 0.007$\\ $0.108 \pm 0.002$\end{tabular}} &
  \multicolumn{1}{c|}{\begin{tabular}[c]{@{}c@{}}$\mathbf{0.668 \pm 0.000}$\\ $\mathbf{3.079 \pm 0.036}$\end{tabular}} &
  \begin{tabular}[c]{@{}c@{}}$0.661 \pm 0.007$\\ $2.222 \pm 0.058$\end{tabular} \\ \cline{6-8} 
\multicolumn{1}{|c|}{\textbf{1}} &
  \multicolumn{1}{c|}{inf-euroroad \cite{Subelj2011}} &
  \multicolumn{1}{c|}{1174} &
  \multicolumn{1}{c|}{1417} &
  infrastructure &
  \multicolumn{1}{c|}{\begin{tabular}[c]{@{}c@{}}$0.502 \pm 0.011$\\ $1.137 \pm 0.412$\end{tabular}} &
  \multicolumn{1}{c|}{\begin{tabular}[c]{@{}c@{}}$\mathbf{0.747 \pm 0.020}$\\ $\mathbf{14.123 \pm 0.830}$\end{tabular}} &
  \begin{tabular}[c]{@{}c@{}}$0.528 \pm 0.008$\\ $9.858 \pm 0.006$\end{tabular} \\ \cline{6-8} 
\multicolumn{1}{|c|}{\textbf{2}} &
  \multicolumn{1}{c|}{bio-celegans \cite{Jeong2000}} &
  \multicolumn{1}{c|}{453} &
  \multicolumn{1}{c|}{2025} &
  biological &
  \multicolumn{1}{c|}{\begin{tabular}[c]{@{}c@{}}$0.562 \pm 0.012$\\ $1.301 \pm 0.018$\end{tabular}} &
  \multicolumn{1}{c|}{\begin{tabular}[c]{@{}c@{}}$0.538 \pm 0.023$\\ $4.256 \pm 0.028$\end{tabular}} &
  \begin{tabular}[c]{@{}c@{}}$\mathbf{0.817 \pm 0.022}$\\ $\mathbf{4.152 \pm 0.013}$\end{tabular} \\ \cline{6-8} 
\multicolumn{1}{|c|}{\textbf{3}} &
  \multicolumn{1}{c|}{in-arenas \cite{arenas_dataset}} &
  \multicolumn{1}{c|}{1133} &
  \multicolumn{1}{c|}{5399} &
  communication &
  \multicolumn{1}{c|}{\begin{tabular}[c]{@{}c@{}}$0.648 \pm 0.011$\\ $1.885 \pm 0.133$\end{tabular}} &
  \multicolumn{1}{c|}{\begin{tabular}[c]{@{}c@{}}$0.807 \pm 0.045$\\ $18.273 \pm 0.050$\end{tabular}} &
  \begin{tabular}[c]{@{}c@{}}$\mathbf{0.884 \pm 0.010}$\\ $\mathbf{18.688 \pm 0.054}$\end{tabular} \\ \cline{6-8} 
\multicolumn{1}{|c|}{\textbf{4}} &
  \multicolumn{1}{c|}{inf-power \cite{watts1998collective}} &
  \multicolumn{1}{c|}{4941} &
  \multicolumn{1}{c|}{6594} &
  infrastructure &
  \multicolumn{1}{c|}{\begin{tabular}[c]{@{}c@{}}$0.506 \pm 0.010$\\ $31.538 \pm 1.721$\end{tabular}} &
  \multicolumn{1}{c|}{\begin{tabular}[c]{@{}c@{}}$\mathbf{0.630 \pm 0.016}$\\ $\mathbf{1013.999 \pm 11.035}$\end{tabular}} &
  \begin{tabular}[c]{@{}c@{}}$0.505 \pm 0.010$\\ $461.960 \pm 1.399$\end{tabular} \\ \cline{6-8} 
\multicolumn{1}{|c|}{\textbf{5}} &
  \multicolumn{1}{c|}{ca-Erdos992 \cite{BATAGELJ2000173}} &
  \multicolumn{1}{c|}{5094} &
  \multicolumn{1}{c|}{7515} &
  collaboration &
  \multicolumn{1}{c|}{\begin{tabular}[c]{@{}c@{}}$0.222 \pm 0.003$\\ $32.595 \pm 1.145$\end{tabular}} &
  \multicolumn{1}{c|}{\begin{tabular}[c]{@{}c@{}}$\mathbf{0.292 \pm 0.003}$\\ $\mathbf{1122.475 \pm 1.156}$\end{tabular}} &
  \begin{tabular}[c]{@{}c@{}}$0.249 \pm 0.003$\\ $498.791 \pm 0.787$\end{tabular} \\ \cline{6-8} 
\multicolumn{1}{|c|}{\textbf{6}} &
  \multicolumn{1}{c|}{ca-GrQc \cite{Grqc}} &
  \multicolumn{1}{c|}{4158} &
  \multicolumn{1}{c|}{13422} &
  collaboration &
  \multicolumn{1}{c|}{\begin{tabular}[c]{@{}c@{}}$0.510 \pm 0.008$\\ $25.477 \pm 0.039$\end{tabular}} &
  \multicolumn{1}{c|}{\begin{tabular}[c]{@{}c@{}}$0.287 \pm 0.007$\\ $700.845 \pm 0.006$\end{tabular}} &
  \begin{tabular}[c]{@{}c@{}}$\mathbf{0.554 \pm 0.013}$\\ $\mathbf{717.982 \pm 0.765}$\end{tabular} \\ \cline{6-8} 
\multicolumn{1}{|c|}{\textbf{7}} &
  \multicolumn{1}{c|}{soc-hamsterster \cite{10.1145/2487788.2488173, dunker2015social}} &
  \multicolumn{1}{c|}{2426} &
  \multicolumn{1}{c|}{16630} &
  social &
  \multicolumn{1}{c|}{\begin{tabular}[c]{@{}c@{}}$0.448 \pm 0.008$\\ $12.621 \pm 0.051$\end{tabular}} &
  \multicolumn{1}{c|}{\begin{tabular}[c]{@{}c@{}}$0.366 \pm 0.003$\\ $185.681 \pm 0.608$\end{tabular}} &
  \begin{tabular}[c]{@{}c@{}}$\mathbf{0.595 \pm 0.005}$\\ $\mathbf{196.021 \pm 0.155}$\end{tabular} \\ \cline{6-8} 
\multicolumn{1}{|c|}{\textbf{8}} &
  \multicolumn{1}{c|}{bio-dmela \cite{dmela}} &
  \multicolumn{1}{c|}{7393} &
  \multicolumn{1}{c|}{25569} &
  biological &
  \multicolumn{1}{c|}{\begin{tabular}[c]{@{}c@{}}$0.437 \pm 0.001$\\ $94.028 \pm 7.124$\end{tabular}} &
  \multicolumn{1}{c|}{\begin{tabular}[c]{@{}c@{}}$0.262 \pm 0.053$\\ $3157.374 \pm 6.624$\end{tabular}} &
  \begin{tabular}[c]{@{}c@{}}$\mathbf{0.584 \pm 0.001}$\\ $\mathbf{2642.664 \pm 1.773}$\end{tabular} \\ \cline{6-8} 
\multicolumn{1}{|c|}{\textbf{9}} &
  \multicolumn{1}{c|}{socfb-Haverford76 \cite{TRAUD20124165}} &
  \multicolumn{1}{c|}{1446} &
  \multicolumn{1}{c|}{59589} &
  social &
  \multicolumn{1}{c|}{\begin{tabular}[c]{@{}c@{}}$0.390 \pm 0.004$\\ $18.178 \pm 0.381$\end{tabular}} &
  \multicolumn{1}{c|}{\begin{tabular}[c]{@{}c@{}}$0.340 \pm 0.006$\\ $42.819 \pm 0.007$\end{tabular}} &
  \begin{tabular}[c]{@{}c@{}}$\mathbf{0.768 \pm 0.016}$\\ $\mathbf{49.867 \pm 0.107}$\end{tabular} \\ \cline{6-8} 
\multicolumn{1}{|c|}{\textbf{10}} &
  \multicolumn{1}{c|}{socfb-Swarthmore42 \cite{TRAUD20124165}} &
  \multicolumn{1}{c|}{1657} &
  \multicolumn{1}{c|}{61049} &
  social &
  \multicolumn{1}{c|}{\begin{tabular}[c]{@{}c@{}}$0.392 \pm 0.012$\\ $25.413 \pm 0.241$\end{tabular}} &
  \multicolumn{1}{c|}{\begin{tabular}[c]{@{}c@{}}$0.646 \pm 0.111$\\ $80.465 \pm 0.097$\end{tabular}} &
  \begin{tabular}[c]{@{}c@{}}$\mathbf{0.954 \pm 0.030}$\\ $\mathbf{96.319 \pm 0.414}$\end{tabular} \\ \cline{6-8} 
\multicolumn{1}{|c|}{\textbf{11}} &
  \multicolumn{1}{c|}{socfb-Bowdoin47 \cite{TRAUD20124165}} &
  \multicolumn{1}{c|}{2250} &
  \multicolumn{1}{c|}{84386} &
  social &
  \multicolumn{1}{c|}{\begin{tabular}[c]{@{}c@{}}$0.429 \pm 0.0021$\\ $48.168 \pm 0.786$\end{tabular}} &
  \multicolumn{1}{c|}{\begin{tabular}[c]{@{}c@{}}$0.175 \pm 0.033$\\ $168.275 \pm 0.248$\end{tabular}} &
  \begin{tabular}[c]{@{}c@{}}$\mathbf{0.823 \pm 0.026}$\\ $\mathbf{202.938 \pm 0.197}$\end{tabular} \\ \cline{6-8} 
\multicolumn{1}{|c|}{\textbf{12}} &
  \multicolumn{1}{c|}{soc-facebook \cite{facebook_nips}} &
  \multicolumn{1}{c|}{4039} &
  \multicolumn{1}{c|}{87352} &
  social &
  \multicolumn{1}{c|}{\begin{tabular}[c]{@{}c@{}}$0.553 \pm 0.016$\\ $95.339 \pm 1.186$\end{tabular}} &
  \multicolumn{1}{c|}{\begin{tabular}[c]{@{}c@{}}$0.158 \pm 0.004$\\ $658.184 \pm 1.422$\end{tabular}} &
  \begin{tabular}[c]{@{}c@{}}$\mathbf{0.739 \pm 0.003}$\\ $\mathbf{732.166 \pm 4.983}$\end{tabular} \\ \cline{6-8} 
\multicolumn{1}{|c|}{\textbf{13}} &
  \multicolumn{1}{c|}{socfb-Hamilton46 \cite{TRAUD20124165}} &
  \multicolumn{1}{c|}{2312} &
  \multicolumn{1}{c|}{96393} &
  social &
  \multicolumn{1}{c|}{\begin{tabular}[c]{@{}c@{}}$0.415 \pm 0.001$\\ $57.185 \pm 0.321$\end{tabular}} &
  \multicolumn{1}{c|}{\begin{tabular}[c]{@{}c@{}}$0.134 \pm 0.007$\\ $181.555 \pm 0.220$\end{tabular}} &
  \begin{tabular}[c]{@{}c@{}}$\mathbf{0.872 \pm 0.002}$\\ $\mathbf{222.257 \pm 0.380}$\end{tabular} \\ \cline{6-8} 
\multicolumn{1}{|c|}{\textbf{14}} &
  \multicolumn{1}{c|}{ca-AstroPh \cite{Grqc}} &
  \multicolumn{1}{c|}{17903} &
  \multicolumn{1}{c|}{195004} &
  collaboration &
  \multicolumn{1}{c|}{\begin{tabular}[c]{@{}c@{}}$0.449 \pm 0.002$\\ $1207.700 \pm 2.354$\end{tabular}} &
  \multicolumn{1}{c|}{\begin{tabular}[c]{@{}c@{}}$0.082 \pm 0.000$\\ $41798.393 \pm 1268.389$\end{tabular}} &
  \begin{tabular}[c]{@{}c@{}}$\mathbf{0.577 \pm 0.001}$\\ $\mathbf{44002.664 \pm 169.846}$\end{tabular} \\ \hline
\multicolumn{1}{|c|}{\textbf{15}} &
  \multicolumn{1}{c|}{Voles* \cite{voles_data}} &
  \multicolumn{1}{c|}{713} &
  \multicolumn{1}{c|}{2322} &
  proximity &
  \multicolumn{1}{c|}{\begin{tabular}[c]{@{}c@{}}0.742\\ 0.521\end{tabular}} &
  \multicolumn{1}{c|}{\begin{tabular}[c]{@{}c@{}}0.978\\ 6.342\end{tabular}} &
  \begin{tabular}[c]{@{}c@{}}\textbf{0.983}\\ \textbf{6.052}\end{tabular} \\ \cline{6-8} 
\multicolumn{1}{|c|}{\textbf{16}} &
  \multicolumn{1}{c|}{HighSchool* \cite{high_school}} &
  \multicolumn{1}{c|}{327} &
  \multicolumn{1}{c|}{5818} &
  proximity &
  \multicolumn{1}{c|}{\begin{tabular}[c]{@{}c@{}}0.281\\ 0.423\end{tabular}} &
  \multicolumn{1}{c|}{\begin{tabular}[c]{@{}c@{}}0.960\\ 3.694\end{tabular}} &
  \begin{tabular}[c]{@{}c@{}}\textbf{1.000}\\ \textbf{2.985}\end{tabular} \\ \cline{6-8} 
\multicolumn{1}{|c|}{\textbf{17}} &
  \multicolumn{1}{c|}{Yeast* \cite{Collins2007, 8013767}} &
  \multicolumn{1}{c|}{1004} &
  \multicolumn{1}{c|}{8323} &
  biological &
  \multicolumn{1}{c|}{\begin{tabular}[c]{@{}c@{}}\textbf{0.655}\\ \textbf{1.797}\end{tabular}} &
  \multicolumn{1}{c|}{\begin{tabular}[c]{@{}c@{}}0.361\\ 13.334\end{tabular}} &
  \begin{tabular}[c]{@{}c@{}}0.33\\ 13.756\end{tabular} \\ \hline
\end{tabular}%
}
\caption{Datasets and accuracy results for numerical experiments on graph alignment using FRTDs and the existing FUGAL \cite{bommakanti2024fugal} algorithm. The best performing algorithm (in terms of accuracy) is labelled in bold for each dataset.
We study the same set of networks previously used in the review of Ref.~\cite{skitsas2023comprehensive}.
Some of these versions of the networks are different from the original data, e.g., they are slightly different sizes.
We could not tell exactly how they are different but use the same versions as Ref.~\cite{skitsas2023comprehensive} for consistency (retrieved from \href{https://github.com/constantinosskitsas/Framework_GraphAlignment}{github.com/constantinosskitsas/Framework\_GraphAlignment}).
Data are also available in the repositories of Refs.~\cite{NetworkRepository, tiago_p_peixoto_2020_7839981, snapnets}.
Indices correspond to the labels in Fig.~\ref{alignment_result1}.
Networks marked with an asterisk correspond to ``real noise'' networks---networks where multiple noisy copies are available and these are aligned to each other rather than to copies with artificial noise. \textit{Voles} and \textit{HighSchool} are temporal proximity networks, where the final network is aligned to a previous variant containing 95\% of edges (we acknowledge the \href{http://www.sociopatterns.org}{SocioPatterns} collaboration for the HighSchool network).
\textit{Yeast} is a protein-protein interaction network for yeast, in this case the ``best'' network is aligned to a noisy variant containing an additional 5\% of low confidence interactions.}
\label{tab:alignmentdata}
\end{table}


\begin{thebibliography}{10}

\bibitem{doi:10.1073/pnas.1401651112}
{\sc Y.~Aflalo, A.~Bronstein, and R.~Kimmel}, {\em On convex relaxation of
  graph isomorphism}, Proceedings of the National Academy of Sciences, 112
  (2015), pp.~2942--2947, \url{https://doi.org/10.1073/pnas.1401651112}.

\bibitem{Monte_Carlo_PageRank}
{\sc K.~Avrachenkov, N.~Litvak, D.~Nemirovsky, and N.~Osipova}, {\em Monte
  {C}arlo methods in {P}age{R}ank computation: When one iteration is
  sufficient}, SIAM Journal on Numerical Analysis, 45 (2007), pp.~890--904,
  \url{https://doi.org/10.1137/050643799}.

\bibitem{BATAGELJ2000173}
{\sc V.~Batagelj and A.~Mrvar}, {\em Some analyses of {E}rd\"os collaboration
  graph}, Social Networks, 22 (2000), pp.~173--186,
  \url{https://doi.org/10.1016/S0378-8733(00)00023-X}.

\bibitem{bhamidi2008mixing}
{\sc S.~Bhamidi, G.~Bresler, and A.~Sly}, {\em Mixing time of exponential
  random graphs}, in 2008 49th Annual IEEE Symposium on Foundations of Computer
  Science, IEEE, 2008, pp.~803--812,
  \url{https://doi.org/10.1109/FOCS.2008.75}.


\bibitem{bommakanti2024fugal}
{\sc A.~Bommakanti, H.~R. Vonteri, K.~Skitsas, S.~Ranu, D.~Mottin, and
  P.~Karras}, {\em {FUGAL}: Feature-fortified unrestricted graph alignment}, in
  The Thirty-eighth Annual Conference on Neural Information Processing Systems,
  2024, \url{https://doi.org/10.52202/079017-0616}.

\bibitem{browet2014algorithms}
{\sc A.~Browet}, {\em Algorithms for community and role detection in networks},
  PhD thesis, Catholic University of Louvain, Louvain-la-Neuve, Belgium, 2014.

\bibitem{cai2018comprehensive}
{\sc H.~Cai, V.~W. Zheng, and K.~C.-C. Chang}, {\em A comprehensive survey of
  graph embedding: Problems, techniques, and applications}, IEEE Transactions
  on Knowledge and Data Engineering, 30 (2018), pp.~1616--1637,
  \url{https://doi.org/10.1109/TKDE.2018.2807452}.

\bibitem{cason2012role}
{\sc T.~P. Cason}, {\em Role extraction in networks}, PhD thesis, Catholic
  University of Louvain, Louvain-la-Neuve, Belgium, 2012.

\bibitem{webkb-link}
CMU World Wide Knowledge Base (WebKB) project, \url{https://www.cs.cmu.edu/afs/cs.cmu.edu/project/theo-11/www/wwkb/}

\bibitem{cookelegans}
{\sc  S.J.~Cook, T.A.~Jarrell, C.A.~Brittin} \textit{et al.}, {\em Whole-animal connectomes of both Caenorhabditis elegans sexes}, Nature, 571 (2019), pp.~63–71, \url{https://doi.org/10.1038/s41586-019-1352-7}. 

\bibitem{Collins2007}
{\sc S.~R. Collins, P.~Kemmeren, X.-C. Zhao, J.~F. Greenblatt, F.~Spencer,
  F.~C. Holstege, J.~S. Weissman, and N.~J. Krogan}, {\em Toward a
  comprehensive atlas of the physical interactome of saccharomyces cerevisiae},
  Molecular {\&} Cellular Proteomics, 6 (2007), pp.~439--450,
  \url{https://doi.org/10.1074/mcp.M600381-MCP200}.

\bibitem{Costa01012007}
{\sc L.~da~F.~Costa, F.~A. Rodrigues, G.~Travieso, and P.~R.~V. Boas}, {\em
  Characterization of complex networks: A survey of measurements}, Advances in
  Physics, 56 (2007), pp.~167--242,
  \url{https://doi.org/10.1080/00018730601170527}.


\bibitem{webkb}
{\sc M.~Craven, D.~DiPasquo, D.~Freitag, A.~McCallum, T.~Mitchell, K.~Nigam, and S.~Slattery}, {\em Learning to construct knowledge bases from the World Wide Web}, Artificial Intelligence, 118 (2000), pp.~69-113, \url{https://doi.org/10.1016/S0004-3702(00)00004-7}


\bibitem{voles_data}
{\sc S.~Davis, B.~Abbasi, S.~Shah, S.~Telfer, and M.~Begon}, {\em Spatial
  analyses of wildlife contact networks}, Journal of The Royal Society
  Interface, 12 (2015), p.~20141004,
  \url{https://doi.org/10.1098/rsif.2014.1004}.

\bibitem{graphwave}
{\sc C.~Donnat, M.~Zitnik, D.~Hallac, and J.~Leskovec}, {\em Learning
  structural node embeddings via diffusion wavelets}, in Proceedings of the
  24th ACM SIGKDD International Conference on Knowledge Discovery \& Data
  Mining, Association for Computing Machinery, 2018, p.~1320–1329,
  \url{https://doi.org/10.1145/3219819.3220025}.

\bibitem{doreian2005generalized}
{\sc P.~Doreian, V.~Batagelj, and A.~Ferligoj}, {\em Generalized
  Blockmodeling}, Cambridge University Press, 2010,
  \url{https://doi.org/10.1017/CBO9780511584176}.

\bibitem{dunker2015social}
{\sc D.~D{\"u}nker and J.~Kunegis}, {\em Social networking by proxy: A case
  study of {C}atster, {D}ogster and {H}amsterster}, arXiv preprint
  arXiv:1501.04527,  (2015).

\bibitem{dwivedi2021graph}
{\sc V.~P. Dwivedi, A.~T. Luu, T.~Laurent, Y.~Bengio, and X.~Bresson}, {\em
  Graph neural networks with learnable structural and positional
  representations}, in The Tenth International Conference on Learning
  Representations, 2022, \url{https://openreview.net/forum?id=wTTjnvGphYj}.

\bibitem{earl2005parallel}
{\sc D.~J. Earl and M.~W. Deem}, {\em Parallel tempering: Theory, applications,
  and new perspectives}, Physical Chemistry Chemical Physics, 7 (2005),
  pp.~3910--3916, \url{https://doi.org/10.1039/B509983H}.

\bibitem{FORTUNATO20161}
{\sc S.~Fortunato and D.~Hric}, {\em Community detection in networks: A user
  guide}, Physics Reports, 659 (2016), pp.~1--44,
  \url{https://doi.org/10.1016/j.physrep.2016.09.002}.

\bibitem{high_school}
{\sc J.~Fournet and A.~Barrat}, {\em Contact patterns among high school
  students}, PLoS ONE, 9 (2014), pp.~1--17,
  \url{https://doi.org/10.1371/journal.pone.0107878}.

\bibitem{fouss2016algorithms}
{\sc F.~Fouss, M.~Saerens, and M.~Shimbo}, {\em Algorithms and models for
  network data and link analysis}, Cambridge University Press, 2016,
  \url{https://doi.org/10.1017/CBO9781316418321}.

\bibitem{Frucht_1949}
{\sc R.~Frucht}, {\em Graphs of degree three with a given abstract group},
  Canadian Journal of Mathematics, 1 (1949), p.~365–378,
  \url{https://doi.org/10.4153/CJM-1949-033-6}.


\bibitem{grover2016node2vec}
{\sc A.~Grover and J.~Leskovec}, {\em node2vec: Scalable feature learning for
  networks}, in Proceedings of the 22nd ACM SIGKDD International conference on
  Knowledge Discovery and Data Mining, 2016, pp.~855--864,
  \url{https://doi.org/10.1145/2939672.2939754}.

\bibitem{arenas_dataset}
{\sc R.~Guimer\`a, L.~Danon, A.~D\'{\i}az-Guilera, F.~Giralt, and A.~Arenas},
  {\em Self-similar community structure in a network of human interactions},
  Physical Review E, 68 (2003), p.~065103,
  \url{https://doi.org/10.1103/PhysRevE.68.065103}.

\bibitem{haemers}
{\sc W.~Haemers}, {\em Are almost all graphs determined by their spectrum?},
  Notices of the South African Mathematical Society, 47 (2016), pp.~42--45.

\bibitem{hamilton2017representation}
{\sc W.~L. Hamilton, R.~Ying, and J.~Leskovec}, {\em Representation learning on
  graphs: Methods and applications}, {IEEE} Data Engineering Bulletin, 40
  (2017), pp.~52--74, \url{https://api.semanticscholar.org/CorpusID:3215337}.

\bibitem{henderson2012rolx}
{\sc K.~Henderson, B.~Gallagher, T.~Eliassi-Rad, H.~Tong, S.~Basu, L.~Akoglu,
  D.~Koutra, C.~Faloutsos, and L.~Li}, {\em Rolx: structural role extraction \&
  mining in large graphs}, in Proceedings of the 18th ACM SIGKDD International
  Conference on Knowledge Discovery and Data Mining, 2012, pp.~1231--1239,
  \url{https://doi.org/10.1145/2339530.2339723}.

\bibitem{henderson2011refex}  
{\sc K.~Henderson, B.~Gallagher, L.~Li, L.~Akoglu, T.~Eliassi-Rad, H.~Tong and C.~Faloutsos}, {\em It's who you know: graph mining using recursive structural features}, in Proceedings of the 17th ACM SIGKDD International Conference on Knowledge Discovery and Data Mining, 2011, pp.~663–671, \url{https://doi.org/10.1145/2020408.2020512}

  

\bibitem{PhysRevE.111.064306}
{\sc E.~Hormann, R.~Lambiotte, and G.~T. Cantwell}, {\em Approximation for
  return time distributions of random walks on sparse networks}, Physical
  Review E, 111 (2025), p.~064306,
  \url{https://doi.org/10.1103/PhysRevE.111.064306}.

\bibitem{Jeong2000}
{\sc H.~Jeong, B.~Tombor, R.~Albert, Z.~N. Oltvai, and A.-L. Barab{\'a}si},
  {\em The large-scale organization of metabolic networks}, Nature, 407 (2000),
  pp.~651--654, \url{https://doi.org/10.1038/35036627}.

\bibitem{exponentialDS}
{\sc I.~Koval and M.~Kwan}, {\em Exponentially many graphs are determined by
  their spectrum}, The Quarterly Journal of Mathematics, 75 (2024),
  pp.~869--899, \url{https://doi.org/10.1093/qmath/haae030}.

\bibitem{kuhn1955hungarian}
{\sc H.~W. Kuhn}, {\em The hungarian method for the assignment problem}, Naval
  research logistics quarterly, 2 (1955), pp.~83--97,
  \url{https://doi.org/10.1002/nav.3800020109}.

\bibitem{10.1145/2487788.2488173}
{\sc J.~Kunegis}, {\em {KONECT}: the {K}oblenz network collection}, in
  Proceedings of the 22nd International Conference on World Wide Web, New York,
  NY, USA, 2013, Association for Computing Machinery, p.~1343–1350,
  \url{https://doi.org/10.1145/2487788.2488173}.

\bibitem{lambiotte2021modularity}
{\sc R.~Lambiotte and M.~T. Schaub}, {\em Modularity and dynamics on complex
  networks}, Cambridge University Press, 2021,
  \url{https://doi.org/10.1017/9781108774116}.

\bibitem{Grqc}
{\sc J.~Leskovec, J.~Kleinberg, and C.~Faloutsos}, {\em Graph evolution:
  Densification and shrinking diameters}, ACM Transactions on Knowledge
  Discovery from Data, 1 (2007), p.~2–es,
  \url{https://doi.org/10.1145/1217299.1217301}.

\bibitem{snapnets}
{\sc J.~Leskovec and A.~Krevl}, {\em {SNAP Datasets}: {Stanford} large network
  dataset collection}.
\newblock \url{http://snap.stanford.edu/data}, June 2014.

\bibitem{facebook_nips}
{\sc J.~Leskovec and J.~Mc{A}uley}, {\em Learning to discover social circles in
  ego networks}, in Advances in Neural Information Processing Systems, vol.~25,
  2012,
  \url{https://proceedings.neurips.cc/paper_files/paper/2012/file/7a614fd06c325499f1680b9896beedeb-Paper.pdf}.

\bibitem{MCKAY1990565}
{\sc B.~D.~McKay and N.~C.~Wormald}, {\em Asymptotic enumeration by degree
  sequence of graphs of high degree}, European Journal of Combinatorics, 11
  (1990), pp.~565--580, \url{https://doi.org/10.1016/S0195-6698(13)80042-X}.

\bibitem{newman2003structure}
{\sc M.~E.~J. Newman}, {\em The structure and function of complex networks},
  SIAM Review, 45 (2003), pp.~167--256,
  \url{https://doi.org/10.1137/S003614450342480}.

\bibitem{newman2006finding}
{\sc M.~E.~J. Newman}, {\em Finding community structure in networks using the
  eigenvectors of matrices}, Physical review E, 74 (2006), p.~036104.

\bibitem{newman2018networks}
{\sc M.~E.~J. Newman}, {\em Networks}, Oxford University Press, 2~ed., 2018,
  \url{https://doi.org/10.1093/oso/9780198805090.001.0001}.

\bibitem{ilprints422}
{\sc L.~Page, S.~Brin, R.~Motwani, and T.~Winograd}, {\em The {P}age{R}ank
  citation ranking: Bringing order to the web.}, Technical Report 1999-66,
  Stanford InfoLab, November 1999.

\bibitem{park2004statistical}
{\sc J.~Park and M.~E.~J. Newman}, {\em Statistical mechanics of networks},
  Physical Review E, 70 (2004), p.~066117,
  \url{https://doi.org/10.1103/PhysRevE.70.066117}.


\bibitem{tiago_p_peixoto_2020_7839981}
{\sc T.~P. Peixoto}, {\em The {N}etzschleuder network catalogue and
  repository}, Aug. 2020, \url{https://doi.org/10.5281/zenodo.7839981}.

\bibitem{penrose2003random}
{\sc M.~Penrose}, {\em Random Geometric Graphs}, Oxford University Press, 2003, \url{https://doi.org/10.1093/acprof:oso/9780198506263.001.0001}.

\bibitem{perozzi2014deepwalk}
{\sc B.~Perozzi, R.~Al-Rfou, and S.~Skiena}, {\em Deepwalk: Online learning of
  social representations}, in Proceedings of the 20th ACM SIGKDD International
  conference on Knowledge Discovery and Data Mining, 2014, pp.~701--710,
  \url{https://doi.org/10.1145/2623330.2623732}.

\bibitem{ribeiro2017struc2vec}
{\sc L.~F.~R. Ribeiro, P.~H.~P. Saverese, and D.~R. Figueiredo}, {\em
  struc2vec: Learning node representations from structural identity}, in
  Proceedings of the 23rd ACM SIGKDD International Conference on Knowledge
  Discovery and Data Mining, 2017, pp.~385--394,
  \url{https://doi.org/10.1145/3097983.3098061}.

\bibitem{directedDP}
{\sc E.~S. Roberts and A.~C.~C. Coolen}, {\em Unbiased degree-preserving
  randomization of directed binary networks}, Physical Review E, 85 (2012),
  p.~046103, \url{https://doi.org/10.1103/PhysRevE.85.046103}.

\bibitem{rossi2014role}
{\sc R.~A. Rossi and N.~K. Ahmed}, {\em Role discovery in networks}, IEEE
  Transactions on Knowledge and Data Engineering, 27 (2014), pp.~1112--1131,
  \url{https://doi.org/10.1109/TKDE.2014.2349913}.

\bibitem{NetworkRepository}
{\sc R.~A. Rossi and N.~K. Ahmed}, {\em The network data repository with
  interactive graph analytics and visualization}, in AAAI, 2015,
  \url{https://networkrepository.com}.

\bibitem{proximity}
{\sc R.~A. Rossi, D.~Jin, S.~Kim, N.~K. Ahmed, D.~Koutra, and J.~B. Lee}, {\em
  On proximity and structural role-based embeddings in networks:
  Misconceptions, techniques, and applications}, ACM Transactions on Knowledge
  Discovery from Data, 14 (2020), \url{https://doi.org/10.1145/3397191}.

\bibitem{Multidynam}
{\sc M.~T. Schaub, J.-C. Delvenne, R.~Lambiotte, and M.~Barahona}, {\em
  Multiscale dynamical embeddings of complex networks}, Physical Review E, 99
  (2019), p.~062308, \url{https://doi.org/10.1103/PhysRevE.99.062308}.

\bibitem{dmela}
{\sc R.~Singh, J.~Xu, and B.~Berger}, {\em Global alignment of multiple protein
  interaction networks with application to functional orthology detection},
  Proceedings of the National Academy of Sciences, 105 (2008),
  pp.~12763--12768, \url{https://doi.org/10.1073/pnas.0806627105}.

\bibitem{skitsas2023comprehensive}
{\sc K.~Skitsas, K.~Orlowski, J.~Hermanns, D.~Mottin, and P.~Karras}, {\em
  Comprehensive evaluation of algorithms for unrestricted graph alignment}, in
  EDBT, 2023, pp.~260--272, \url{https://doi.org/10.48786/edbt.2023.21}.

\bibitem{snijders2002markov}
{\sc T.~A.~B. Snijders}, {\em Markov chain {M}onte {C}arlo estimation of
  exponential random graph models}, Journal of Social Structure, 3 (2002),
  pp.~1--40.

\bibitem{Subelj2011}
{\sc L.~{\v{S}}ubelj and M.~Bajec}, {\em Robust network community detection
  using balanced propagation}, The European Physical Journal B, 81 (2011),
  pp.~353--362, \url{https://doi.org/10.1140/epjb/e2011-10979-2}.

\bibitem{PhysRevLett.57.2607}
{\sc R.~H. Swendsen and J.-S. Wang}, {\em Replica monte carlo simulation of
  spin-glasses}, Physical Review Letters, 57 (1986), pp.~2607--2609,
  \url{https://doi.org/10.1103/PhysRevLett.57.2607}.

\bibitem{TANG20251}
{\sc R.~Tang, Z.~Yong, S.~Jiang, X.~Chen, Y.~Liu, Y.-C. Zhang, G.-Q. Sun, and
  W.~Wang}, {\em Network alignment}, Physics Reports, 1107 (2025), pp.~1--45,
  \url{https://doi.org/10.1016/j.physrep.2024.11.006}.

\bibitem{TRAUD20124165}
{\sc A.~L. Traud, P.~J. Mucha, and M.~A. Porter}, {\em Social structure of
  Facebook networks}, Physica A: Statistical Mechanics and its Applications,
  391 (2012), pp.~4165--4180,
  \url{https://doi.org/10.1016/j.physa.2011.12.021}.

\bibitem{8013767}
{\sc V.~Vijayan and T.~Milenković}, {\em Multiple network alignment via
  {M}ulti{MAGNA}++}, IEEE/ACM Transactions on Computational Biology and
  Bioinformatics, 15 (2018), pp.~1669--1682,
  \url{https://doi.org/10.1109/TCBB.2017.2740381}.

\bibitem{faq}
{\sc J.~T. Vogelstein, J.~M. Conroy, V.~Lyzinski, L.~J. Podrazik, S.~G.
  Kratzer, E.~T. Harley, D.~E. Fishkind, R.~J. Vogelstein, and C.~E. Priebe},
  {\em Fast approximate quadratic programming for graph matching}, PLOS ONE, 10
  (2015), pp.~1--17, \url{https://doi.org/10.1371/journal.pone.0121002}.

\bibitem{von_luxburg_tutorial_2007}
{\sc U.~Von~Luxburg}, {\em A tutorial on spectral clustering}, Statistics and
  Computing, 17 (2007), pp.~395--416,
  \url{https://doi.org/10.1007/s11222-007-9033-z}.

\bibitem{Wang03042025}
{\sc W.~Wang and W.~Wang}, {\em Haemers’ conjecture: An algorithmic
  perspective}, Experimental Mathematics, 34 (2025), pp.~147--161,
  \url{https://doi.org/10.1080/10586458.2024.2337229}.

\bibitem{watts1998collective}
{\sc D.~J. Watts and S.~H. Strogatz}, {\em Collective dynamics of
  ‘small-world’ networks}, Nature, 393 (1998), pp.~440--442,
  \url{https://doi.org/10.1038/30918}.

\bibitem{white1986structure}
{\sc J.~G. White, E.~Southgate, J.~N. Thomson, and S.~Brenner}, {\em The
  structure of the nervous system of the nematode {C}aenorhabditis elegans: the
  mind of a worm}, Philosophical Transactions of the Royal Society B, 314
  (1986), pp.~1--340, \url{https://doi.org/10.1098/rstb.1986.0056}.

\bibitem{zachary1977information}
{\sc W.~W. Zachary}, {\em An information flow model for conflict and fission in
  small groups}, Journal of Anthropological Research, 33 (1977), pp.~452--473,
  \url{http://www.jstor.org/stable/3629752}.

\bibitem{zhang2018retgk}
{\sc Z.~Zhang, M.~Wang, Y.~Xiang, Y.~Huang, and A.~Nehorai}, {\em Ret{GK}:
  Graph kernels based on return probabilities of random walks}, in Proceedings
  of the 32nd International Conference on Neural Information Processing
  Systems, vol.~31, 2018,
  \url{https://proceedings.neurips.cc/paper_files/paper/2018/file/7f16109f1619fd7a733daf5a84c708c1-Paper.pdf}.

\end{thebibliography}
\end{document}